\documentclass[letterpaper,twocolumn,10pt]{ilass}
\def\thepage{}

\usepackage{amssymb}
\usepackage{amsmath}

\usepackage{ifpdf}

\ifpdf
\usepackage[pdftex]{graphicx}
\else
\usepackage{graphicx}
\fi

\usepackage{subfigure}
\usepackage{subfigmat}
\usepackage{color}

\usepackage{booktabs}

\usepackage{epstopdf}
\usepackage[capitalize]{cleveref}

\usepackage[english]{babel}
\usepackage{blindtext}
\usepackage{amsmath,amsfonts,stmaryrd,amssymb,theorem}

\graphicspath{{./figures/}}

\title{
    \vspace{-0.25in} 
    {\small \em AIAA SciTech Forum, Grapevine, Texas, 9-13 January 2017, 55th AIAA Aerospace Sciences Meeting} 
    \newline \newline
    \large {\bf MVP-Workshop Contribution: Modeling of Volvo bluff body flame experiment}
}

\author{
    \large
    Hao Wu, Peter C. Ma, Yu Lv and Matthias Ihme\footnote{Corresponding Author: mihme@stanford.edu}\\
    Department of Mechanical Engineering\\
    Stanford University\\
    Stanford, CA 94305\\
}

\date{\normalsize  \centerline{\bf Abstract} \vspace{0.05in}
\begin{minipage}{6.5in}
\normalsize
The Volvo burner features the canonical configuration of a bluff-body stabilized premixed flame. This configuration was studied experimentally under the Volvo Flygmotor AB program. Two cases are considered in this study: a non-reacting case with an inlet flow speed of 16.6 m/s and a reacting case with equilibrium ratio of 0.65 and inflow speed of 17.3 m/s. The characteristic vortex shedding in the wake behind the bluff body is present in the non-reacting case, while two oscillation modes are intermittently present in the reacting case. A series of large-eddy simulations are performed on this configuration using two solvers, one using a high-resolution finite-volume (FV) scheme and the other featuring a high-order discontinuous-Galerkin (DG) discretization. The FV calculations are conducted on hexahedral meshes with three different resolution (4mm, 2mm, and 1mm). The DG calculations are performed using two different polynomial orders on the same tetrahedral mesh. For the non-reacting cases, good agreement with respect to the experimental data is achieved by both solvers at high numerical resolution. The reacting cases are calculated using a two-step global mechanism in combination with the thickened-flame model. Reasonable agreement with experiments is obtained by both solvers at higher resolution. Models for combustion-turbulence interaction are necessary for the reacting case as it contains the length scale of the flame, which is smaller than the grid resolution in all calculations. The impact of such models on the flame stability and flow/flame dynamics is the subject of future research. 
\end{minipage} \vspace{-0.25in}
}

\begin{document}

\ifpdf
\DeclareGraphicsExtensions{.pdf, .jpg}
\else
\DeclareGraphicsExtensions{.eps, .jpg}
\fi

\maketitle

\clearpage
\pagenumbering{arabic}
\setcounter{page}{2}

\section{Introduction}
The pressing need of propulsion systems with higher efficiency drives the development of lean combustion. However, this combustion mode faces challenges in terms of flame stabilization~\cite{shanbhogue2009lean}. The inherent unsteadiness related to flame stabilization calls for high-fidelity numerical modeling techniques such as Large Eddy Simulations (LES). Although, it has been widely accepted that significant improvement can be achieved in the predictability of turbulent reacting flow simulations through the adoption of LES from RANS~\cite{pitsch2006large}, many issues around the usage of LES remain unclear, largely due to the vast variety of methods that may be applied by LES. 

For reacting flow LES, the modeling choices that may impact the simulation results include the numerical schemes, modeling of sub-grid scale (SGS) turbulence, representation of combustion chemistry, and modeling of turbulence-chemistry interaction. The solution of LES can be affected by the interplay between numerical schemes and turbulence SGS models, because in common practices of LES, the effects of the turbulence SGS model and the grid resolution occur at the same scale~\cite{bose2010grid}. The choice of combustion chemistry models can also affect the LES solution as they are usually constructed by simplifying the detailed chemistry process for the purpose of reducing computational cost~\cite{pope2013}. The process of simplification introduces many assumptions, whose impact and limitations under different circumstances are often not fully understood even by expert users~\cite{wu2015pareto,wu2015fidelity,wu2016compliance}. In addition, the interaction between the combustion process and the turbulent fluid motion at the sub-grid-scale level also needs to be modeled. This is a particularly challenging issue if the Karlovitz number is below order unity, in which case the small-scale structure is affected by such interaction.

To better understand the effectiveness and predictability of LES in face of the aforementioned challenges, this paper carries out a series of simulations using the configuration of the bluff-body stabilized premixed flame. This is a canonical reacting flow configuration that was experimentally investigated under the Volvo Flygmotor AB program~\cite{sjunnesson1991lda,sjunnesson1991validation,ab1992alaa}. The simplistic set-up resembles many aspects of practical combustion devises and presents a combination of interesting physical phenomena, such as combustion instabilities and lean blowout (LBO). In addition, extensive measurements are made available by the experimentalists, which include statistics of both velocity and scalars in the form of first and second moments.

The Volvo configuration has also been extensively studied numerically by other researchers through LES over the past two decades. Among the more recent efforts, the investigation on the effect of numerical methods include the series of simulations performed by Cocks et al.~\cite{cocks2013reacting, cocks2015impact}, in which the results produced by four FV-based solvers using different numerics are compared against each other and experimental data, and the work of Li et al.~\cite{li2016large}, which studies the effect of reflecting and non-reflecting inlets on combustion dynamics. The self-excited combustion instabilities in this configuration were studied by Ghani et al.~\cite{ghani2015longitudinal}. The effects of chemical kinetics are studied by Sardeshmukh et al.~\cite{sardeshmukh2016impact}, Fureby~\cite{fureby2007comparison}, and Manickam et al.~\cite{manickam2012large}. The impacts of more sophisticated treatment of turbulence-combustion interaction are also investigated using LES/PDF method~\cite{kim2014effects} as well as Eulerian stochastic field method~\cite{jones2015large}.

Among the many aforementioned issues, this study primarily focuses on the effect of numerical methods in this study. Specifically, we will investigate the behavior and performance  of a high-resolution finite-volume (FV) scheme, which is the prevailing approach in the LES community, as well as a particular implementation of a discontinuous Galerkin (DG) scheme, which is particularly suitable for achieving higher-order accuracy for convection-dominant flows. Both methods will be tested and validated by performing a series of reacting and non-reacting calculations at different levels of numerical resolution, while using similar approaches in the modeling of combustion chemistry and the turbulence-chemistry interaction.

The rest of the paper is organized as follows: Section~\ref{SEC_CASE} presents the description of the case set-up, including the experimental configuration and the computational domain. Section~\ref{SEC_NUM} describes the finite-volume and the discontinuous-Galerkin discretization schemes employed by the two solvers. The combustion model and the computational grids are also briefly summarized. The results of both reacting and non-reacting simulations produced by the two solvers are presented and discussed in Section~\ref{SEC_RES}. The paper is concluded in Section~\ref{SEC_CONC}.

\section{Case Description} \label{SEC_CASE}

\subsection{Volvo test rig}
\begin{figure*}[!htb!]
    \centering
    \includegraphics[width=0.95\textwidth]{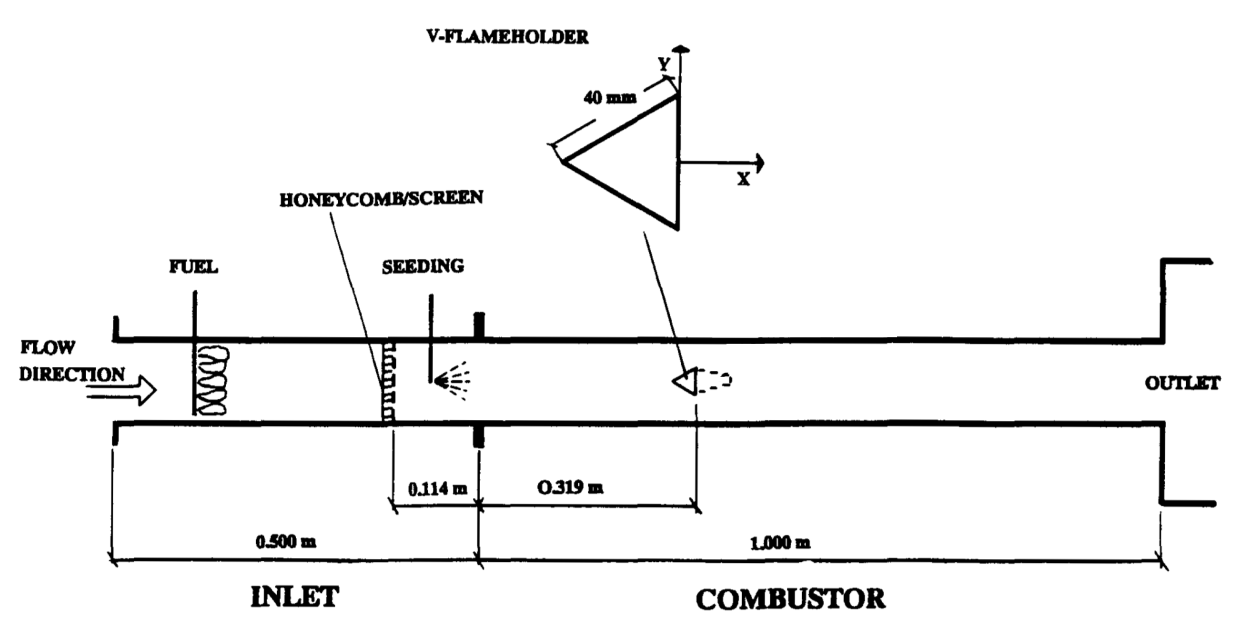}
    \caption{Schematic of the Volvo test rig (adapted from Sjunnesson~et~al.~\cite{ab1992alaa}). \label{fig:testrig}}
\end{figure*}

A schematic of the Volvo test rig is shown in \cref{fig:testrig}. The configuration consists of a 1.5~m long rectangular duct which is 0.24~m wide by 0.12~m high with a choked air inlet. The apparatus is divided into the inlet and combustor sections, which are 0.5~m and 1.0~m in length, respectively. The inlet section is responsible for fueling, seeding, and flow conditioning. A multi-orifice critical flow injector is utilized for the injection of gaseous propane upstream of the inlet section for the reacting cases. Honeycombs located in the inlet section are used to control the turbulence level and mixture homogeneity. A 0.04~m equilateral triangle is mounted 0.682~m upstream from the exit spanning the width of the duct, and acts as the bluff-body flameholder. The flow exits the rectangular duct through a circular outlet.

Experimental measurements available for the Volvo test rig includes velocity data from Laser Doppler Anemometry (LDA)~\cite{sjunnesson1991lda}, and temperature and species data from Coherent Anti-Stokes Raman Scattering (CARS) technique~\cite{ab1992alaa}. Mean and root mean square (RMS) velocity data are available for axial profiles along the center-line and transverse profiles across the height of the combustor section at several axial locations, downstream the flameholder. 

The operating conditions under consideration are summarized in \cref{tab:conditions}. An air mass flow rate of 0.6~kg/s is considered. The Reynolds number of 48,000 is defined using the characteristic length of the bluff body ($D$~=~0.04~m) and bulk inlet velocity and viscosity at the temperature of 288~K. For the reacting case the equivalence ratio is 0.65, which gives an adiabatic flame temperature of 1784~K.

\begin{table*}[!htb!]
    \centering
    \begin{tabular}{c c c c c c}
        \toprule 
        Case & Re & $U_\text{bulk}$ [m/s] & $\phi$ & $T_\text{in}$ [K] & $T_\text{ad}$ [K]\\
        \midrule
        Non-reacting & 48,000 & 16.6 & 0.0 & 288 & - \\
        Reacting & 48,000 & 17.3 & 0.65 & 288 & 1784\\
        \bottomrule
    \end{tabular}
    \caption{Operating conditions. \label{tab:conditions}}
\end{table*}

\subsection{Computational domain}
\begin{figure*}[!htb!]
    \centering
    \includegraphics[width=0.9\textwidth]{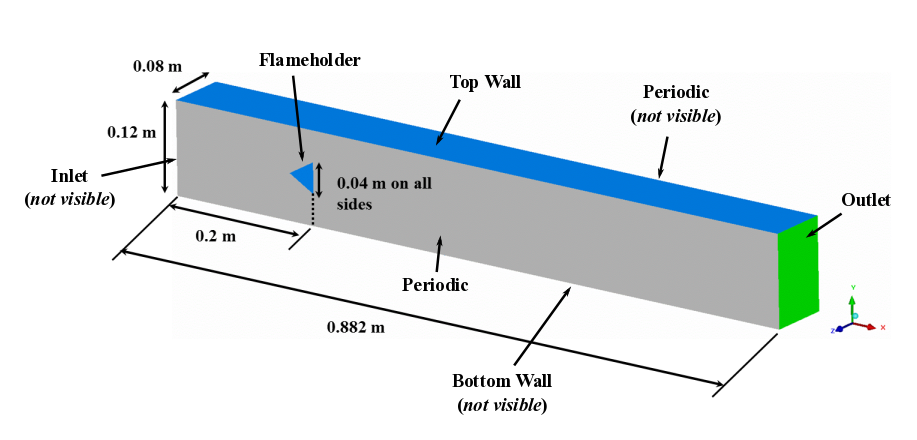}
    \caption{Computational domain and boundary conditions~\cite{mvpws2016}. \label{fig:domain}}
\end{figure*}

The computational domain considered in this study is shown in \cref{fig:domain}, with a depth twice the width of the bluff body, corresponding to one third of the channel in the experiment. The full combustor section downstream of the flameholder is considered. The bluff body is placed 0.2~m from the inlet. Periodic boundary conditions are applied in the spanwise direction. A fixed mass flow rate of 0.2083 kg/s is employed at the inlet through characteristic boundary conditions applied by velocity, temperature and species. No turbulence profile is included for the inlet velocity and a plug-flow profile is adopted. Pressure of 1~bar is specified through characteristic boundary conditions at the outlet. No-slip adiabatic wall boundary conditions are employed at the flameholder and top and bottom walls, following the recommendation of the workshop.

\section{Numerical Setup} \label{SEC_NUM}

\subsection{Governing equations}
The fully compressible Navier-Stokes equations are considered as the governing equations in this study. The Favre-averaged conservation equations of mass, momentum, total energy, and species, are written as follows:
\begin{subequations}
    \label{eqn:governingEqn}
    \begin{align}
        &\frac{\partial \bar{\rho}}{\partial t} + \frac{\partial \bar{\rho} \widetilde{u}_j}{\partial x_j} = 0\,,\\
        &\begin{aligned}[t]
            &\frac{\partial \bar{\rho} \widetilde{u}_i}{\partial t} + \frac{\bar{\rho} \widetilde{u}_i \widetilde{u}_j}{\partial x_j} = -\frac{\partial \bar{p}}{\partial x_i} \\
            &+ \frac{\partial}{\partial x_j} \left[ (\widetilde{\mu} + \mu_t) \left( \frac{\partial \widetilde{u}_i}{\partial x_j} + \frac{\partial \widetilde{u}_j}{\partial x_i} - \frac{2}{3} \delta_{ij} \frac{\partial \widetilde{u}_k}{\partial x_k} \right) \right] \,,
        \end{aligned}\label{eqn:momentumequation} \\
        &\begin{aligned}[t]
            &\frac{\partial  \bar{\rho} \widetilde{E}}{\partial t} + \frac{\bar{\rho} \widetilde{u}_j \widetilde{E}}{\partial x_j} =\\
            &\frac{\partial}{\partial x_j} \left[ \left(\widetilde{\frac{\lambda}{c_p}} + \frac{\mu_t}{\text{Pr}_t} \right) \frac{\partial \widetilde{h}}{\partial x_j} - \widetilde{u}_j \bar{p} + \widetilde{u}_i (\bar{\tau}_{ij} + \bar{\tau}^R_{ij}) \right] \\
            &+ \frac{\partial}{\partial x_j} \left[ \sum_{k = 1}^N \left(\bar{\rho}\widetilde{D}_k - \widetilde{\frac{\lambda}{c_p}} \right) \widetilde{h}_k \frac{\partial \widetilde{Y}_k}{\partial x_j} \right] \,,
        \end{aligned} \label{eqn:energyequation} \\
        &\frac{\partial \bar{\rho} \widetilde{Y}_k}{\partial t} + \frac{\bar{\rho} \widetilde{u}_j \widetilde{Y}_k}{\partial x_j} = \frac{\partial}{\partial x_j} \left[ \left( \bar{\rho}\widetilde{D}_k + \frac{\mu_t}{\text{Sc}_t} \frac{\partial \widetilde{Y}_k}{\partial x_j} \right) \right] + \bar{\dot{\omega}}_k\,,
    \end{align}
\end{subequations}
where $u_i$ is the i-th component of the velocity vector, $E$ is the total energy including the chemical energy, $C$ is the progress variable, $\mu$ and $\mu_t$ are the laminar and turbulent viscosity, $\lambda$ is the thermal conductivity, $D_k$ is the diffusion coefficient for the species $k$, $\dot{\omega}_k$ is the source term for species $k$, $\tau_{ij}$ and $\tau_{ij}^R$ are the viscous and subgrid-scale stresses, Pr$_t$ is the turbulent Prandtl number, and Sc$_t$ is the turbulent Schmidt number. An appropriate subgrid-scale model is needed for the computation of the turbulent viscosity. The system is closed by the ideal-gas law as the equation of state. The sub-grid terms associated with the equation of state are neglected in this study.

\subsection{Finite-volume discretization}
The massively paralleled {\it CharLES$^{\,x}$}, developed at Center for Turbulence Research, Stanford University, is used in this study as the finite-volume solver. {\it CharLES$^{\,x}$} has been applied to studies of several different flow problems including aeroacoustics~\cite{nichols2012large}, supercritical flows~\cite{hickey2013large}, supersonic combustion~\cite{larsson2015incipient}, and aerodynamic flows~\cite{bodart2013large}. A brief summary of the numerical schemes is provided here, and for more details the interested readers are referred to the work by Ham~et~al.~\cite{ham2004energy} and Khalighi~et~al.~\cite{khalighi2011unstructured}.

\begin{figure*}[!htb!]
    \centering
    \subfigure[4 mm]{
        \includegraphics[width=0.9\textwidth]{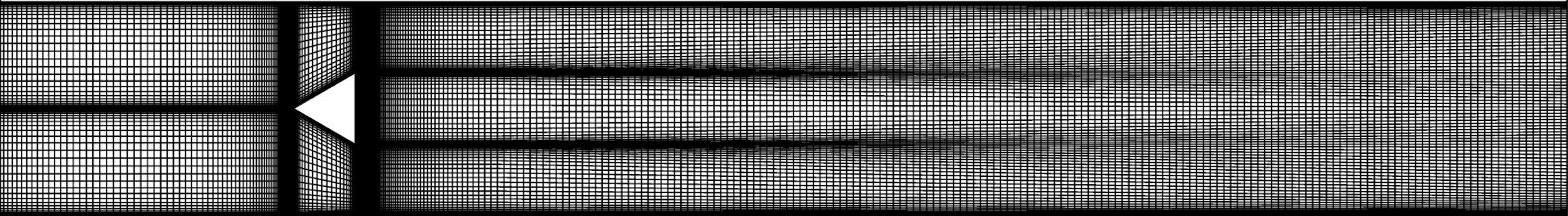}
    }
    \subfigure[2 mm]{
        \includegraphics[width=0.9\textwidth]{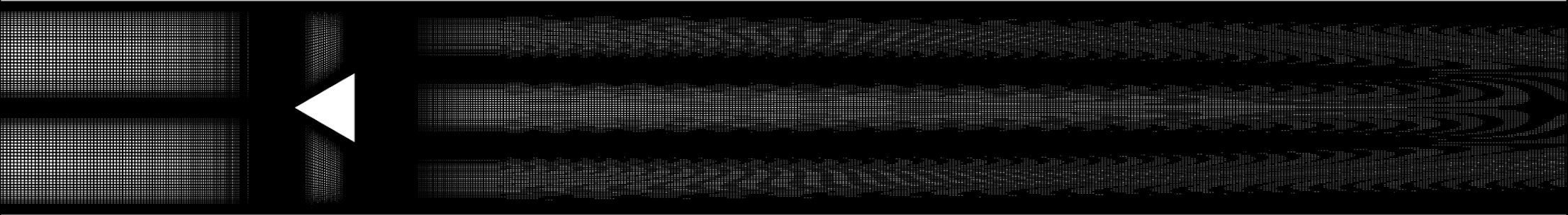}            
    }
    \subfigure[4 mm (top half flameholder)]{
        \includegraphics[trim={0 30.5cm 0 0},width=0.42\textwidth,clip]{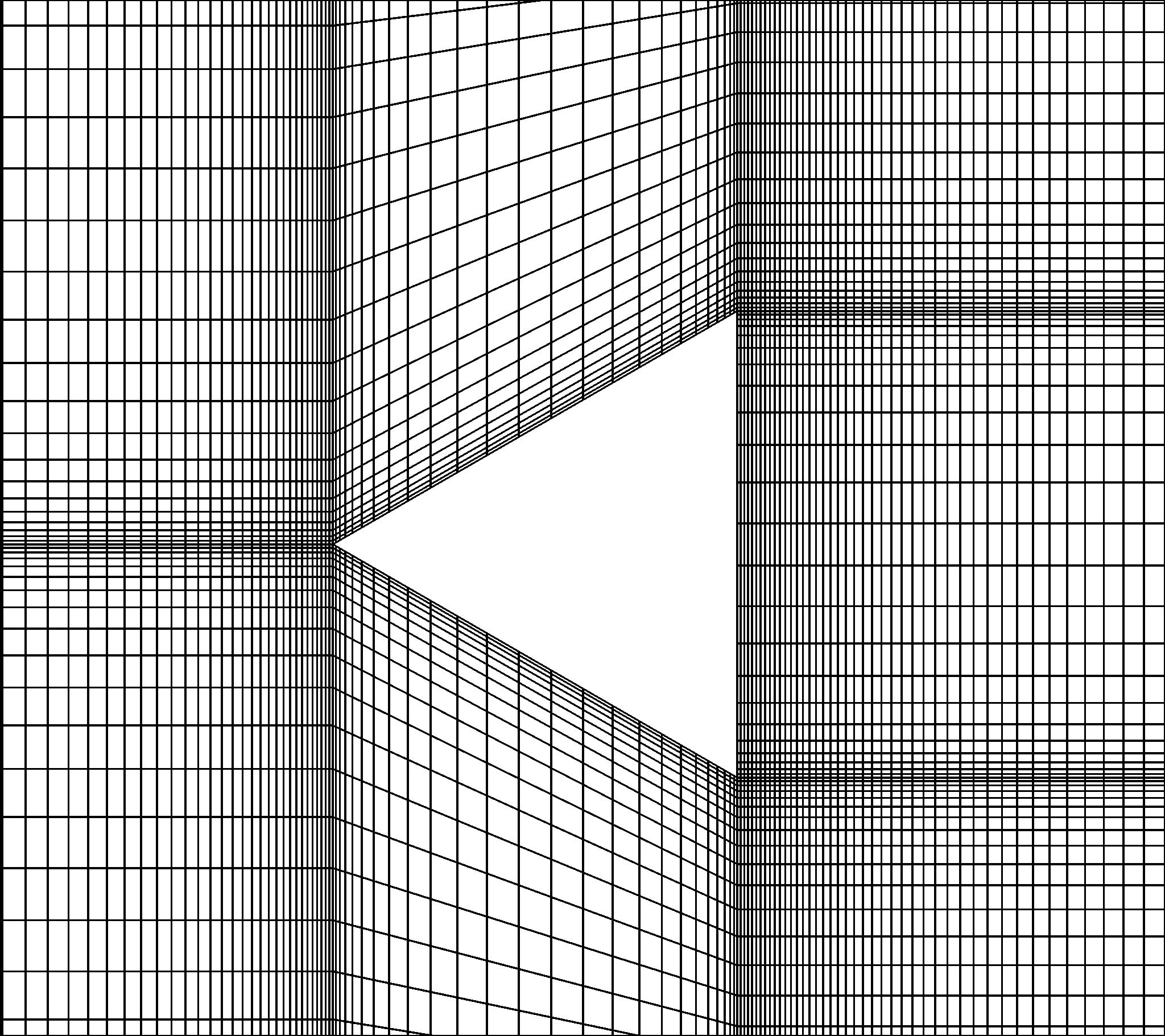}
    } \quad
    \subfigure[2 mm (top half flameholder)]{
        \includegraphics[trim={0 30.5cm 0 0},width=0.42\textwidth,clip]{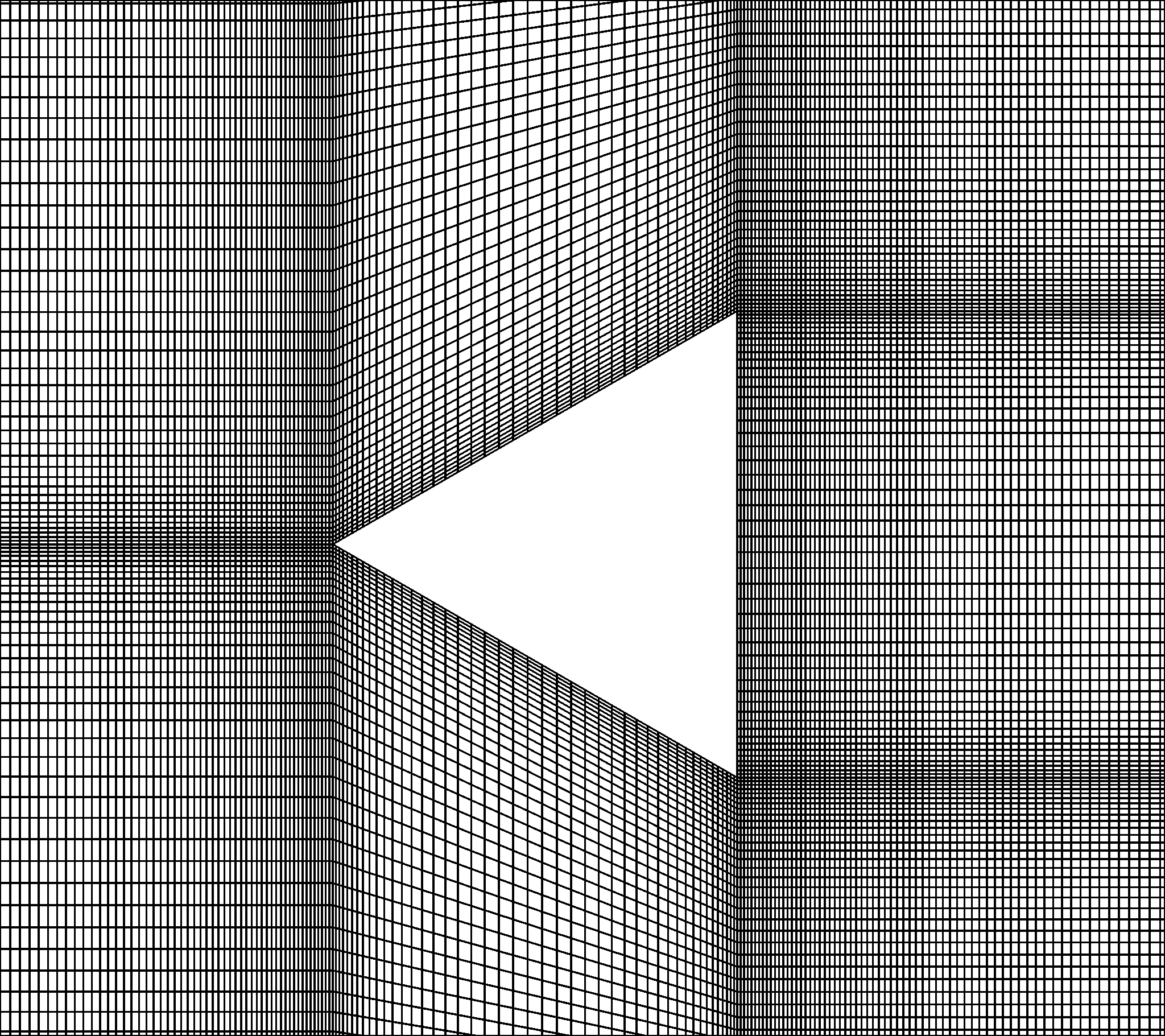}
    }
    \caption{Computational grids for the FV solver. \label{fig:meshfv}}
\end{figure*}

A control-volume based FV approach is utilized for the discretization of the system of \cref{eqn:governingEqn}:
\begin{equation}
    \frac{\partial U}{\partial t} V_{cv} + \sum_f F^e A_f = \sum_f F^v A_f + S\Delta_{cv}\,,
\end{equation}
where $U$ is the vector of conserved variables, $F^e$ is the face-normal Euler flux vector, $F^v$ is the face-normal viscous flux vector, which corresponds to the r.h.s of \cref{eqn:governingEqn}, $S$ is the source term vector, $\Delta_{cv}$ is the volume of the control volume, and $A_f$ is the face area. A strong stability preserving 3rd-order Runge-Kutta (SSP-RK3) scheme \cite{gottlieb2001strong} is used for time advancement. {\it CharLES$^{\,x}$} was developed based on a reconstruction-based FV scheme. Polynomials with maximum third-order accuracy are used to reconstruct the left- and right-biased face centroid values of the flow variables, and a blending between a central flux and a Riemann flux is computed based on the local grid quality. This flux computation procedure yields formally second-order accuracy and has maximum fourth-order accuracy on a perfectly uniform Cartesian mesh, without numerical dissipation. For computations of shock-related problems and cases where large gradients of flow variables are present, {\it CharLES$^{\,x}$} utilizes a sensor-based hybrid Central-ENO scheme to minimize the numerical dissipation while stabilizing the simulation. For regions where shocks and large gradients are present, a second-order ENO reconstruction is used at the left- and right-biased face values, followed by an HLLC Riemann flux computation. A number of different hybrid sensors/switches are available. For this study, a \`relative solution" (RS) sensor is used which is based on the solution and reconstruction values of density~\cite{khalighi2011unstructured,ma2014supercritical}. An entropy-stable flux correction technique~\cite{ma2017entropy} is used to ensure the physical realizability of the numerical solution and to dampen non-linear instabilities.

Following Pope's criterion~\cite{pope2004ten}, Cocks~et~al.~\cite{cocks2013reacting} showed that a resolution of 3~mm is required so that 80\% of the turbulent kinetic energy can be fully resolved by the LES, although the criterion may not be sufficient for reacting cases. Based on this argument, three grids with increasing resolutions of 4~mm, 2~mm, and 1~mm are generated. The 4~mm and 2~mm grids are shown in \cref{fig:meshfv} along with zoomed-in regions near the flameholder. The grids are clustered near the wall, with a minimum wall spacing of 0.3~mm, and the mesh is nearly isotropic in the rest of the domain. A uniform grid is employed in the spanwise direction. The three grids amount to a total of 0.54, 4.17, and 24.6 million cells, respectively. 

A Vreman SGS model~\cite{Vreman2004} is used for the sug-grid terms. The sub-grid scale eddy viscosity is calculated as
\begin{equation}
    \begin{aligned}
    &\mu_t = \bar{\rho} C \Delta_{cv}^{2/3} \\
    &\left( \frac{\beta_{11}\beta_{22}-\beta_{12}^2+\beta_{11}\beta_{33}-\beta_{13}^2+\beta_{22}\beta_{33}-\beta_{23}^2}{\alpha_{ij}\alpha_{ij}} \right)^{1/2}\,,
    \end{aligned}
\end{equation}
where $\alpha_{ij} = \partial \widetilde{u}_j / \partial x_i$, $\beta_{ij} = \alpha_{mi}\alpha_{mj}$, and $\Delta_{cv}$ is the local cell volume. The model constant $C$ is set to a value of 0.07 in this study.

The combustion process in the FV simulations is described by a two-step kinetic scheme as mentioned in the workshop document~\cite{mvpws2016}, which is adapted from Ghani et~al.~\cite{ghani2015longitudinal}. The laminar flame speed and the adiabatic flame temperature obtained by this mechanism are listed in Table~\ref{tab:flamespeed} in comparison with those from the GRI 3.0 mechanism~\cite{smith1999gri}. The difference in the flame speed is $11 \%$ and the flame temperature is virtually identical. 

\begin{table*}[!htb!]
    \centering
    \begin{tabular}{c c c c c c}
        \toprule 
        Mechanism & $S_\text{l}$ [m/s] & $T_\text{b}$ [K]\\
        \midrule
        2-step & 0.194 &  1805.8 \\
        GRI30 & 0.218 & 1806.6 \\
        \bottomrule
    \end{tabular}
    \caption{Laminar flame speed and the adiabatic flame temperature obtained by the two-step mechanism for the FV simulations in comparison to GRI30~\cite{smith1999gri}. \label{tab:flamespeed}}
\end{table*}

The thickness of the flame is estimated to be 0.6 mm, which cannot be resolved by the meshes used in this study. Therefore, a dynamic thickened flame model~\cite{colin2000thickened,legier2000dynamically} is used to describe the flame turbulence interactions with the model of Charlette et al.~\cite{charlette2002power} for the sub-grid efficiency. 
 
\subsection{Discontinuous-Galerkin discretization}
For notational convenience, the governing equations of \cref{eqn:governingEqn} are written in vector form as:
\begin{equation}
    \label{NS_VEC_FORM}
    \frac{\partial}{\partial t} \mathsf{U} + \nabla \cdot \mathsf{F} = \nabla \cdot \mathsf{Q} + \mathsf{S}\;,
\end{equation}
in which $\mathsf{U}$, $\mathsf{F}$, $\mathsf{Q}$, and $\mathsf{S}$ refer to the solution vector, the convective flux, the viscous flux, and the source term, respectively. To discretize Eq.~(\ref{NS_VEC_FORM}) with a variational approach, a broken space is used as the test space
\begin{equation}
\mathcal{V}_h^p = \{\phi\in L^2(\Omega): \phi_e \equiv \phi|_{\Omega_e} \in \mathcal{P}_p, \forall \Omega_e \in \Omega \}\;,
\end{equation}
in which a polynomial space with order $p$ is defined on each individual element $\Omega_e$ resulting from the domain partition. The test function $\phi$ is then multiplied to Eq.~(\ref{NS_VEC_FORM}) and integrated to derive the weak form. The left-hand-side of Eq.~(\ref{NS_VEC_FORM}) is the classical Euler equations, and the corresponding weak forms can be written as:
\begin{equation}
    \label{DG_DISC_CONVECTION}
    \begin{split}
        &\int_{\Omega} \phi (\partial_t \mathsf{U} + \nabla \cdot  \mathsf{F} )d x \\
        &= \sum_e \left(\int_{\Omega_e} (\phi_e \partial_t \mathsf{U} - \nabla \phi_e \cdot \mathsf{F} ) dx + \int_{\partial \Omega_e} \phi_e^+ \mathsf{F}\cdot n ds\right) \;\\
        & \approx \sum_e \left(\int_{\Omega_e} \phi_e \phi_e^T  d_t  \widetilde{U}_{e,i}(t)  dx - \int_{\Omega_e} \nabla\phi_e \cdot  F(U_e) dx \right.\\
        &+\left. \int_{\partial \Omega_e} \phi_e^+ \hat{F} ds \right)\;,
    \end{split}
\end{equation}
in which the superscript ``+" refers to interior quantity on the element edge and the discretized solution $U$ (together with $U_e$ and $ \widetilde{U}_{e,i}$), following the Galerkin method, takes the form
\begin{subequations}
    \begin{align}
        \mathsf{U} \simeq U &= \oplus_{e=1}^{N_e} U_e \;, \\
        U_e(t, x) &= \sum_{i=1}^{N_p} \widetilde{U}_{e,i}(t) \phi_{e,i}(x) \;.
    \end{align}
\end{subequations}
To complete the discretization, the Riemann flux $\widehat{F}$ in Eq.~(\ref{DG_DISC_CONVECTION}) needs to be specified. We consider the preconditioned Roe scheme~\cite{turkel1997assessment} for the current study. 

Since $\mathsf{Q}$ involves multi-entries, here we only consider a general form for each of them to simplify the explanation. $\mathsf{Q}_i$ denotes the diffusion flux of Navier-Stokes equations for the $i^{\text{th}}$ solution variable. This can be further represented using a linearization $\mathsf{Q}_i = \sum_j \mathsf{D}_{ij} \circ \nabla \mathsf{U}_j $, where $\circ$ refers to Hadamard product. Because the discretization is subject to the distributive property of addition, the problem can be further simplified by discretizing $ \mathsf{D}_{ij} \circ \nabla \mathsf{U}_j \equiv \mathsf{Q}_{ij}$ taking the following form
\small
\begin{equation*}
    \tiny
    \begin{split}
        &\sum_e \int_{\Omega_e} \phi \nabla \cdot \mathsf{Q}_{ij} dx \\
        &= \sum_e\left(- \int_{\Omega_e} (\mathsf{D}_{ij} \circ \nabla \mathsf{U}_j)  \cdot \nabla \phi_e dx + \int_{\partial \Omega_e} \phi_e \mathsf{Q}_{ij} \cdot n ds   \right)\;\\
        & =  \sum_e\left(\int_{\Omega_e} \mathsf{U}_j  \nabla \cdot ( \mathsf{D}_{ij} \circ \nabla \phi_e)   dx  - \int_{\partial \Omega_e} \mathsf{U}_j   ( \mathsf{D}_{ij} \circ \nabla \phi_e)  \cdot n ds + \int_{\partial \Omega_e} \phi_e \mathsf{Q}_{ij} \cdot n ds   \right)\;\\
        & \approx \sum_e\left(\int_{\Omega_e} U_j  \nabla \cdot ( {D}_{ij} \circ \nabla \phi_e)   dx  - \int_{\partial \Omega_e} \widehat{U}_j   ( {D}_{ij} \circ \nabla \phi_e)^+  \cdot n ds + \int_{\partial \Omega_e} \phi_e^+ \widehat{{Q}}_{ij}  ds   \right)\;\\
        & = \sum_e \left( - \int_{\Omega_e} \nabla \phi_e  \cdot (D_{ij} \circ \nabla U_j) dx + \int_{\partial \Omega_e}  (U_j^+ - \widehat{U}_j ) ( {D}_{ij} \circ \nabla \phi_e)^+ \cdot n ds +  \int_{\partial \Omega_e} \phi_e^+ \widehat{{Q}}_{ij}ds\right)\;.\\
    \end{split}
\end{equation*}
\normalsize
With this, the discretization of $\nabla \cdot \mathsf{Q}_{i}$ can be written as
\small
\begin{equation}
    \label{DG_DISC_DIFFUSION}
    \begin{split}
        &\sum_e \int_{\Omega_e} \phi \nabla \cdot \mathsf{Q}_i dx \approx\\
        &-\sum_e \int_{\Omega_e}  \nabla \phi_e \cdot ([D_{i}]^{T} \nabla U)dx \\
        &+\sum_e \int_{\partial \Omega_e}  ([D_{i}^+]^{T} (U^+ - \widehat{U}) \circ  \nabla \phi_e)  \cdot nds \\
        &+\sum_e \int_{\partial \Omega_e}  \phi_e^+ \widehat{Q}_i ds \;,
    \end{split}
\end{equation}
\normalsize
where the diffusion Jacobian $[D_i]$ has $N_s$ rows composed of $D_{ij}$ with $j = i \cdots N_s$. The three terms on the right-hand side account for interior diffusion, dual consistency and inter-element diffusion. $\widehat{U}$ and $\widehat{Q}$ can be selected in various ways, resulting different diffusion-discretization schemes. In this study, we consider the standard interior penalty (SIP) method~\cite{arnold2002unified} and $\widehat{U}$ is chosen to be $\{U\}$ where $\{\cdot\} =( (\cdot)^+ + (\cdot)^-)/2$, then we have
\begin{align}
    \label{DISCR_DIFF_FLUX}
    \widehat{Q}_i &= \{[D_i]^T \nabla U  \} + \sigma_{\text{SIP}} \left(\max\limits_{x \in \partial \Omega_e^\pm} \mu (x) \right) \llbracket U  \rrbracket / h\,,
\end{align}
where $\llbracket \cdot \rrbracket = (\cdot)^+n^+ + (\cdot)^- n^-$ and $\sigma$ refers to constant parameters to ensure numerical stability. As for $\sigma_{\text{SIP}}$, we use the suggested values of~\cite{shahbazi2005explicit}. The stabilization term for $\widehat{Q}_i$ in Eq.~(\ref{DISCR_DIFF_FLUX}), having a consistent form of that in the Riemann solvers, always behaves as a dissipation term.

\begin{figure*}[!htb!]
    \centering
    \includegraphics[trim={1.48cm 0.7cm 0.5cm 0.5cm},width=0.75\textwidth,clip]{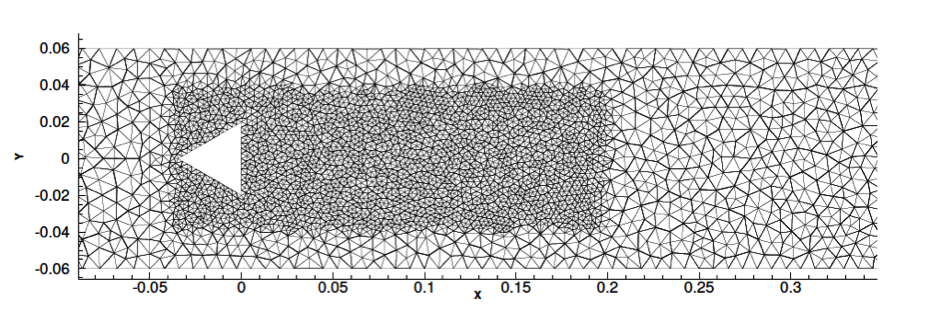}
    \caption{Computational grids for the DG solver. \label{fig:meshdg}}
\end{figure*}

For the present study, the DG-based prediction is carried out on an unstructured mesh that is discretized with about 25,000 tetrahedral elements, as shown in \cref{fig:meshdg}. Two DG schemes, namely DGP1 and DGP2, are considered, which represent element-wise solutions using linear and quadratic polynomials, respectively. DGP1 and DGP2 schemes consists of four and ten polynomial coefficients to be solved in each element. This corresponds to 1.0 and 2.5~millions total degrees of freedom for DGP1 and DGP2 predictions, respectively. A standard five-stage fourth-order Runge-Kutta scheme is employed for time integration. The prediction of reacting flows requires consideration of variable thermal properties (i.e., heat capacity as a function of temperature). Without proper treatment, the standard numerical procedure might produce spurious pressure oscillation. To avoid this issue, a Double-Flux model proposed by Billet and Ryan~\cite{BILLET_RYAN_JCP2011} is used~\cite{LV_IHME_JCP_2014}.

For the present study, no explicit turbulence model is employed, which is instead accommodated using inherent numerical dissipation~\cite{ma2015discontinuous}. An attempt of using a similar strategy to the heat and mass transfer in the reacting simulations shows that insufficient numerical dissipation. As a result, a thickened flame model is used with a constant thickening factor of $20$ for DGP1 and $10$ for DGP2. The SGS efficiency is modeled using a constant efficiency factor of $5$.

\section{Results and  Discussions} \label{SEC_RES}
In this section, the reacting and non-reacting simulations are conducted by two solvers, one using the FV discretization and the other using the DG discretization. The results are presented in comparison with the experimental data.

\subsection{Non-reacting simulations: finite-volume discretization}
\begin{figure*}[!htb!]
    \centering
    \includegraphics[trim={2cm 26cm 2cm 18.5cm},width=0.95\textwidth,clip]{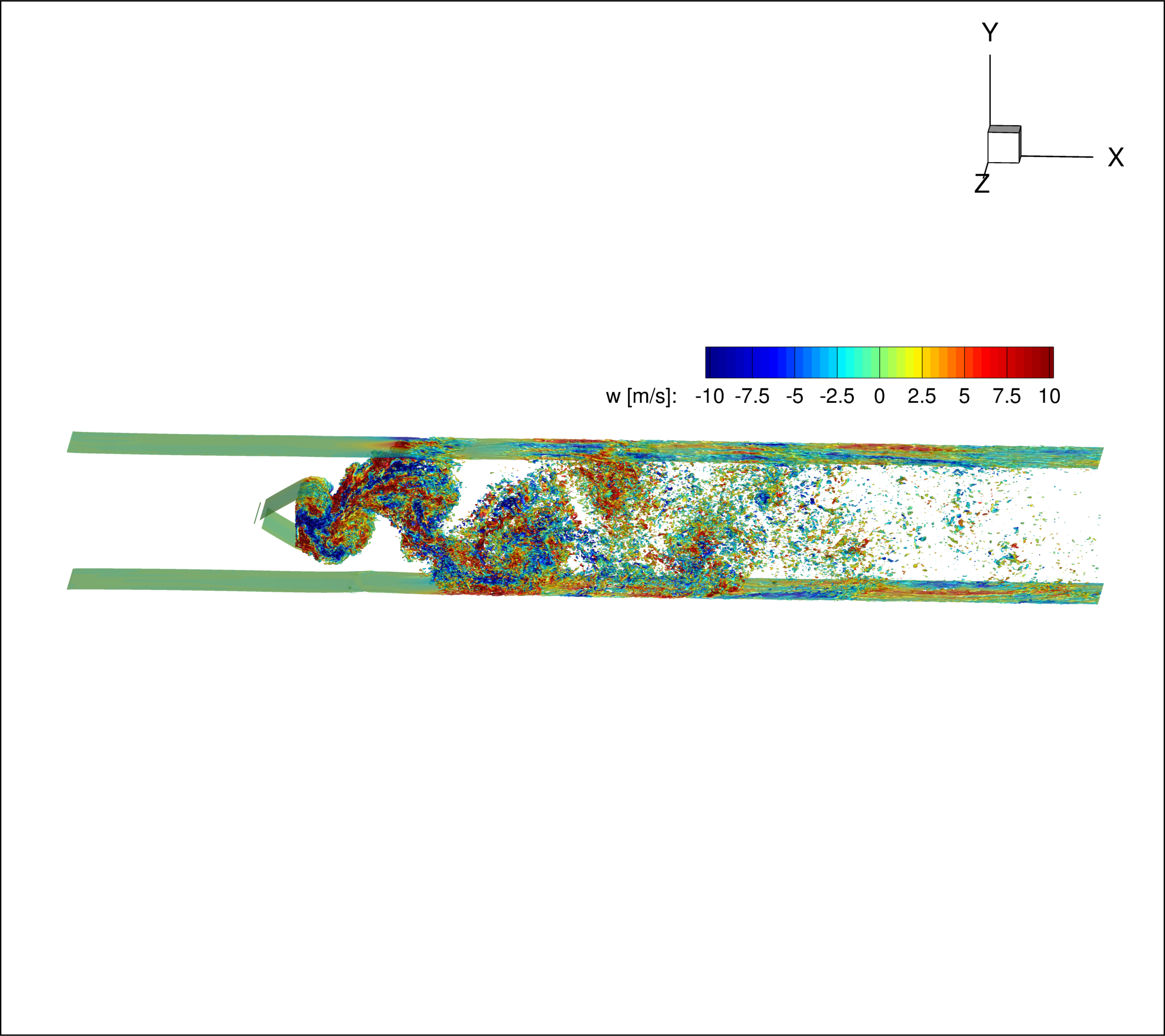}
    \caption{Iso-surfaces of vorticity magnitude of 4000~s$^{-1}$ colored by the spanwise velocity on the 1~mm grid for the non-reacting case from the FV solver. \label{fig:vorticity}}
\end{figure*}

The non-reacting case is simulated with the FV solver using three different resolutions, namely 4~mm, 2~mm, and 1~mm. For the sake of validating the sub-grid scale models utilized and establishing grid convergence study. \Cref{fig:vorticity} shows the vortical structures exhibited downstream of the bluff body from the simulation using the 1~mm grid. Iso-surfaces of vorticity magnitude of 4000~s$^{-1}$ colored by the spanwise velocity are displayed. It can be seen that the von-Karman type shedding is generated behind the flameholder due to the instabilities produced by the flow separations. The large vortices then break up into smaller vortices and the flow becomes more turbulent further downstream. It is also important to observe the interactions between the vortices and the wall. It has been pointed out by Cocks~et~al.~\cite{cocks2013reacting} that the wall vortices play a major role inestablishing three-dimensional vortex breakdown further downstream of the domain.

\begin{figure*}[!htb!]
    \centering
    \includegraphics[trim={30 0 40 0},width=0.48\textwidth,clip]{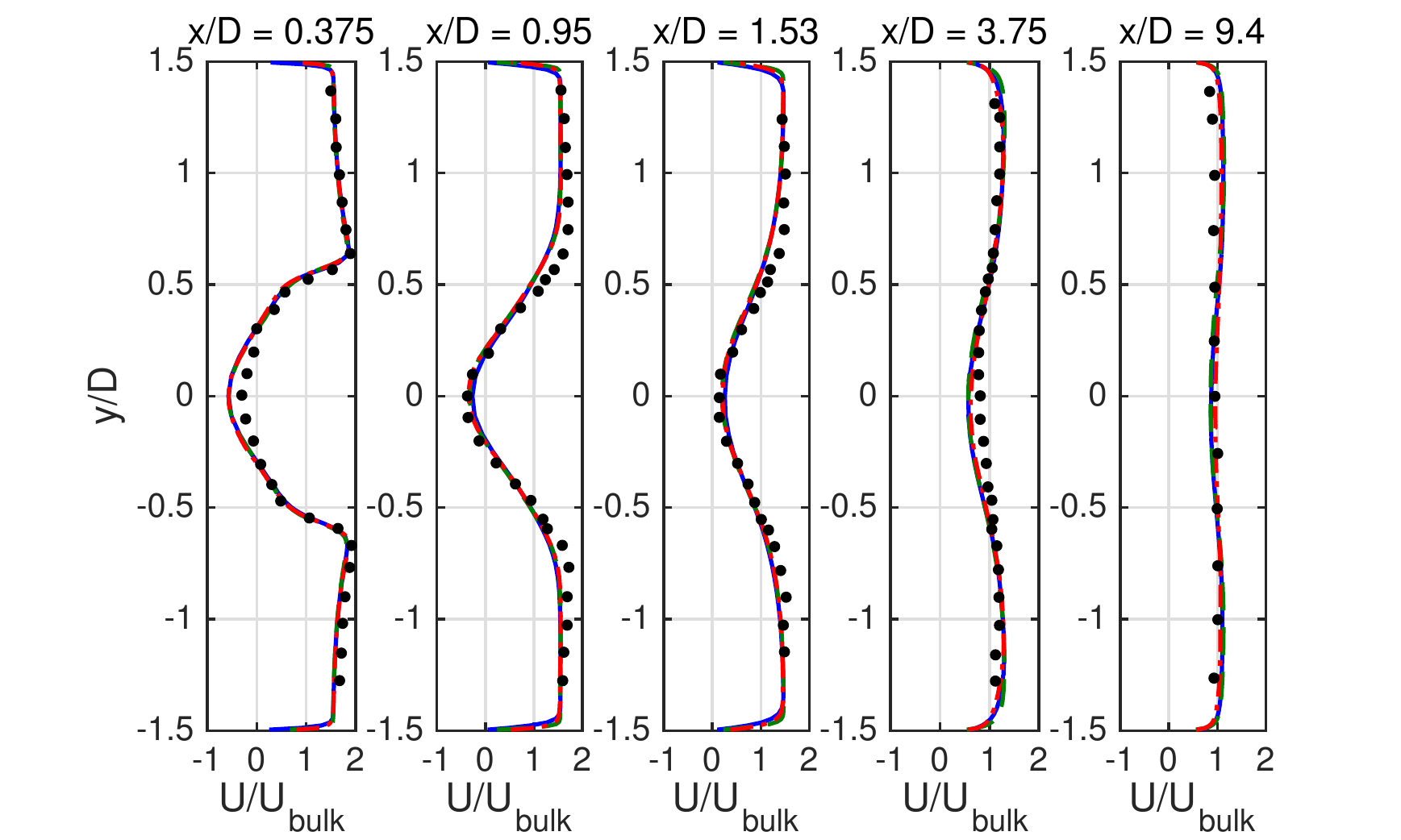}
    \includegraphics[trim={30 0 40 0},width=0.48\textwidth,clip]{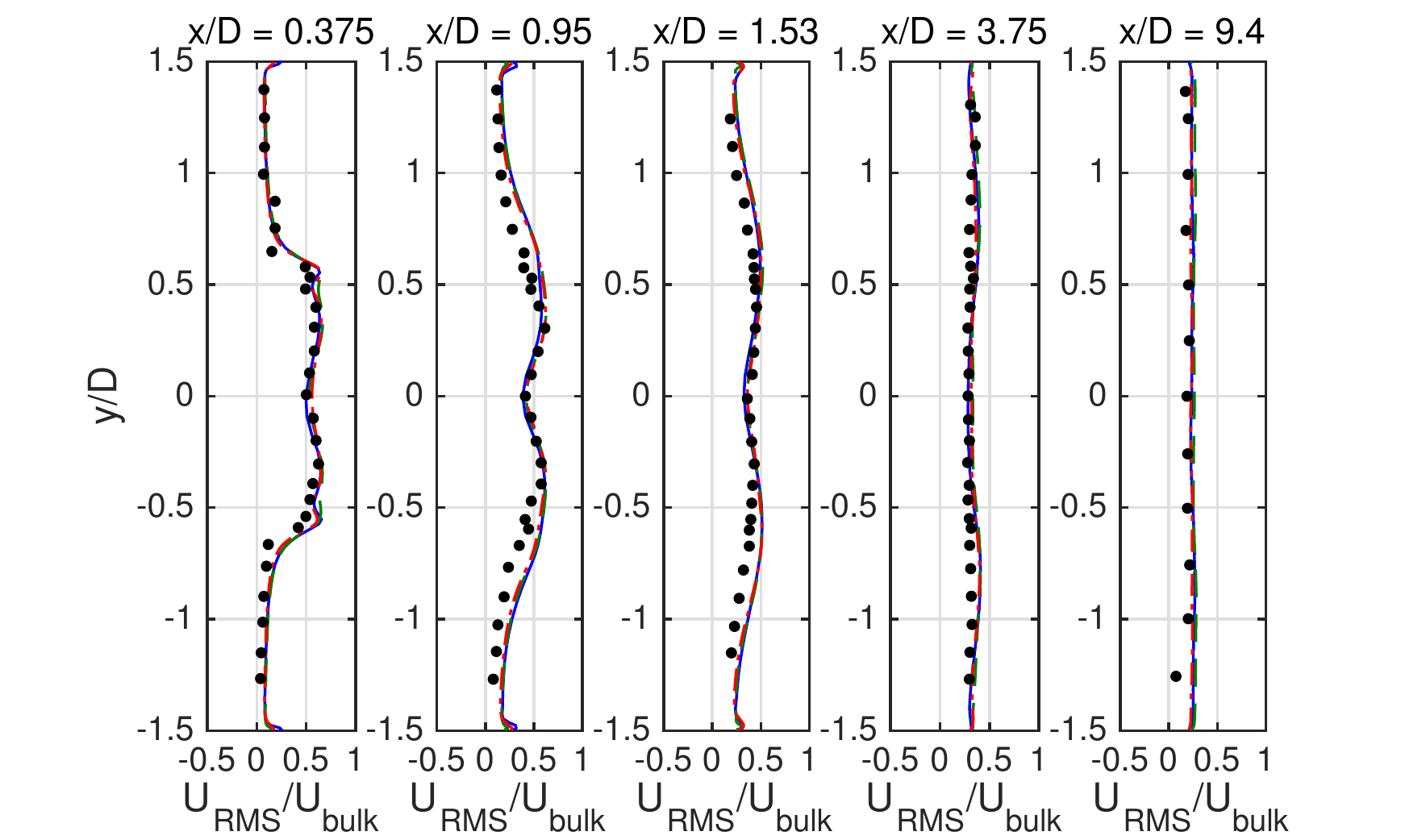}
    \caption{Normalized mean (left) and RMS (right) axial velocity profiles on three grids (solid line~\textendash\,~4~mm, dashed line~\textendash\,~2~mm, dash-dotted line~\textendash\,~1~mm) in comparison with measurements (dots), at several axial locations for the non-reacting case from the FV solver. \label{fig:velocityu}}
\end{figure*}

\begin{figure*}[!htb!]
    \centering
    \includegraphics[trim={30 0 40 0},width=0.48\textwidth,clip]{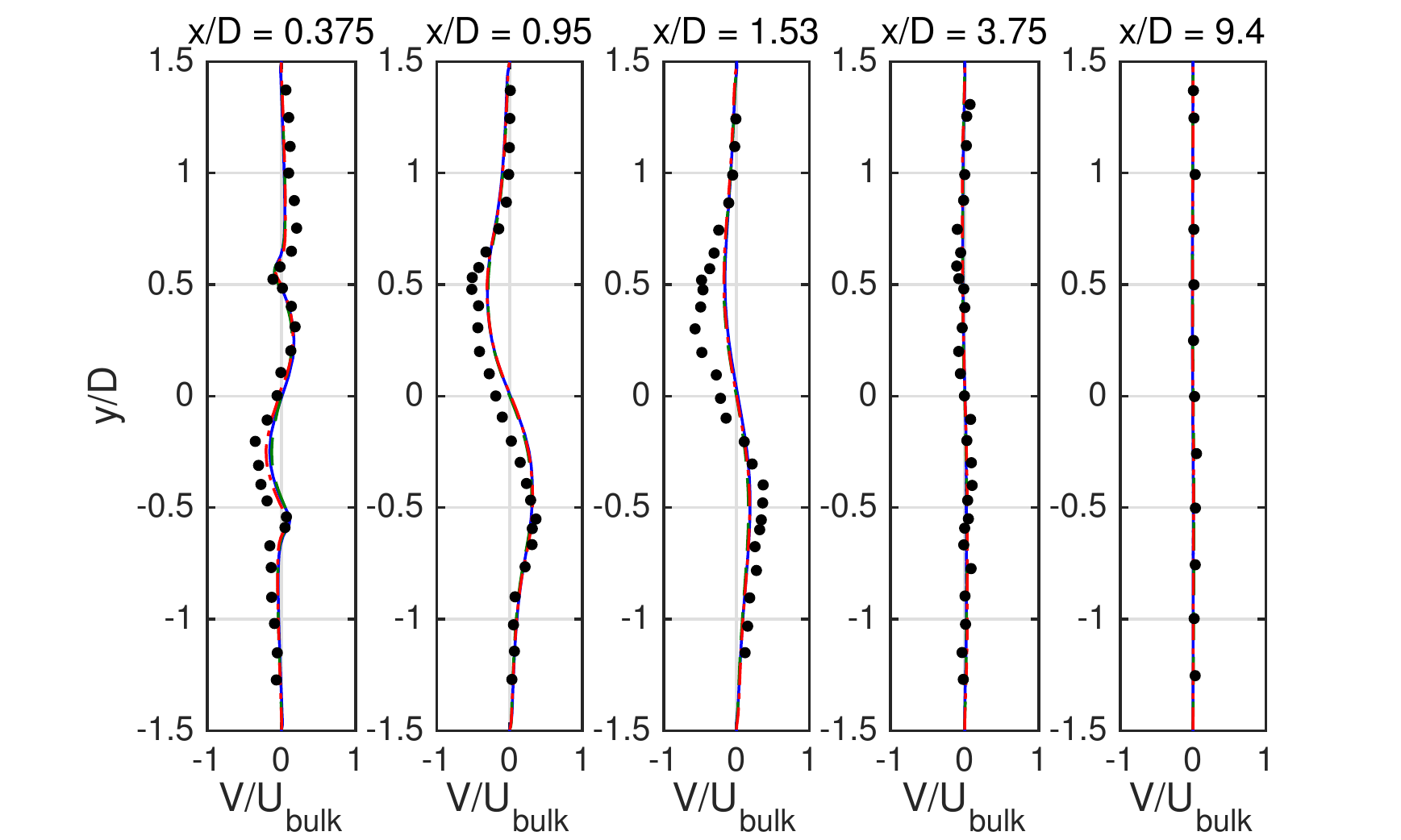}
    \includegraphics[trim={30 0 40 0},width=0.48\textwidth,clip]{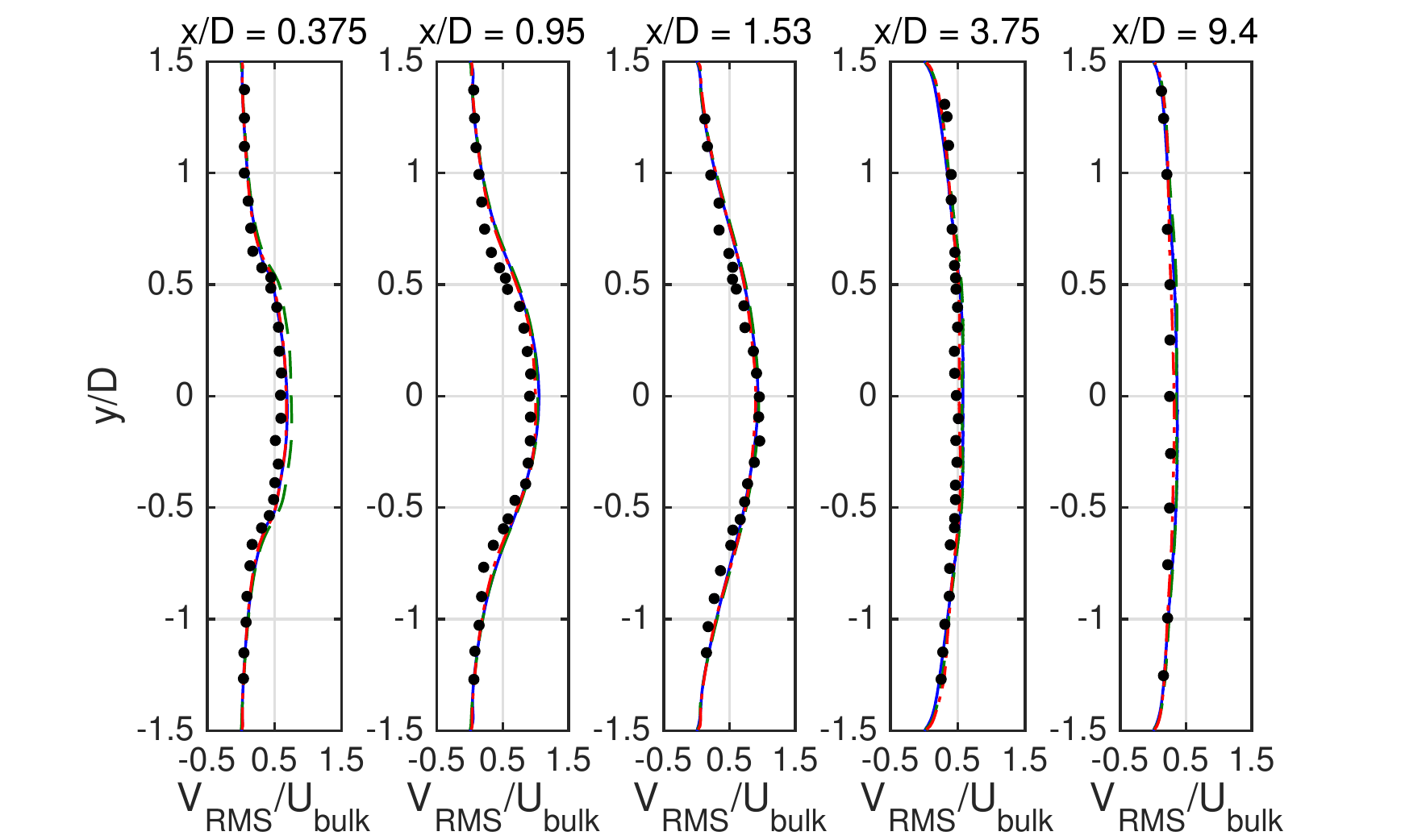}
    \caption{Normalized mean (left) and RMS (right) transverse velocity profiles on three grids (solid line~\textendash\,~4~mm, dashed line~\textendash\,~2~mm, dash-dotted line~\textendash\,~1~mm) in comparison with measurements (dots), at several axial locations for the non-reacting case from the FV solver. \label{fig:velocityv}}
\end{figure*}

\begin{figure*}[!htb!]
    \centering
    \includegraphics[trim={30 0 40 0},width=0.48\textwidth,clip]{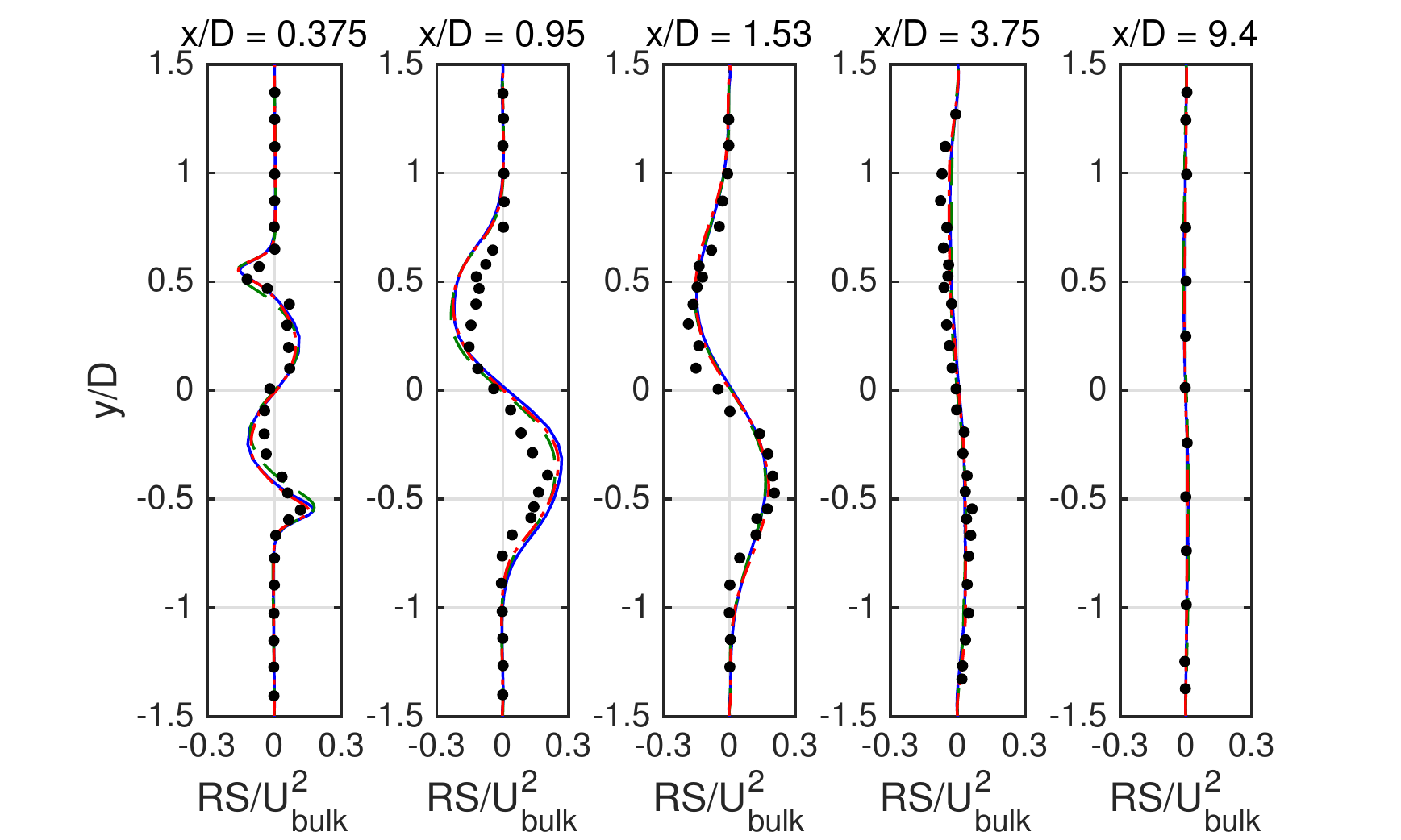}
    \caption{Normalized mean Reynolds stress profiles on three grids (solid line~\textendash\,~4~mm, dashed line~\textendash\,~2~mm, dash-dotted line~\textendash\,~1~mm) in comparison with measurements (dots), at several axial locations for the non-reacting case from the FV solver. \label{fig:reynolds}}
\end{figure*}

\Crefrange{fig:velocityu}{fig:reynolds} show the results of the velocity statistical profiles for the three different grids used. The simulations are averaged for about 300~ms, which corresponds to about six flow-through times over the entire computational domain. Five different axial locations are considered for comparison with the experimental measurements. Mean and root-mean-square (RMS) axial and transverse velocity profiles are shown in \cref{fig:velocityu,fig:velocityv}, and the Reynolds stress profiles are shown in \cref{fig:reynolds}. It can be seen that grid convergence is achieved already with the 2~mm grid for both the first and second moment statistics, and the simulation with the 1~mm grid confirms the convergence. Excellent agreement can be observed for all the quantities compared with the measurements, except for the transverse velocity at the axial location $x/D = 1.53$, where the experiments have larger magnitude.

\begin{figure*}[!htb!]
    \centering
    \subfigure[Axial velocity]{
        \includegraphics[trim={0 0 0 0},width=0.48\textwidth,clip]{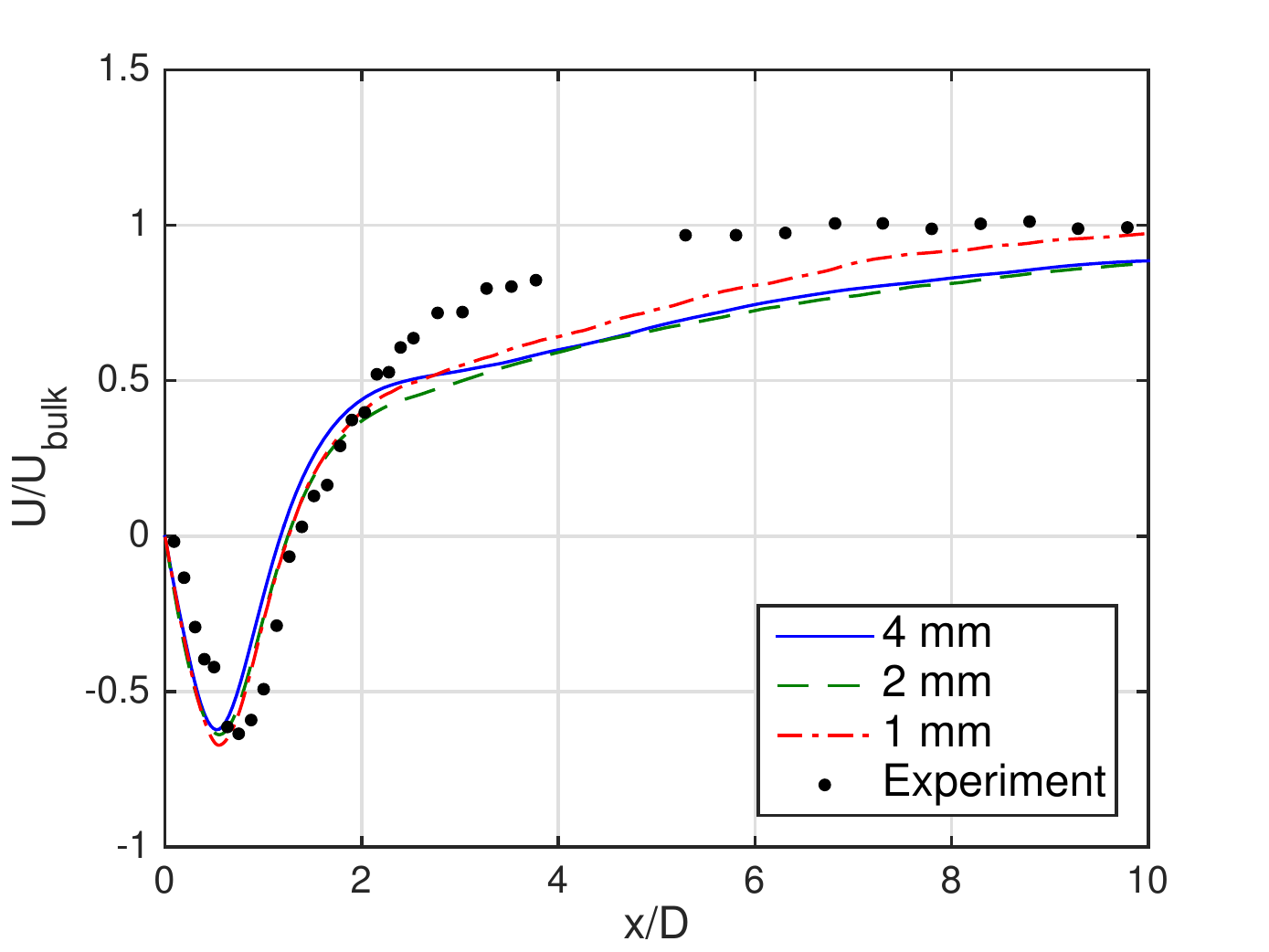}
    }
    \subfigure[Anisotropy]{
        \includegraphics[trim={0 0 0 0},width=0.48\textwidth,clip]{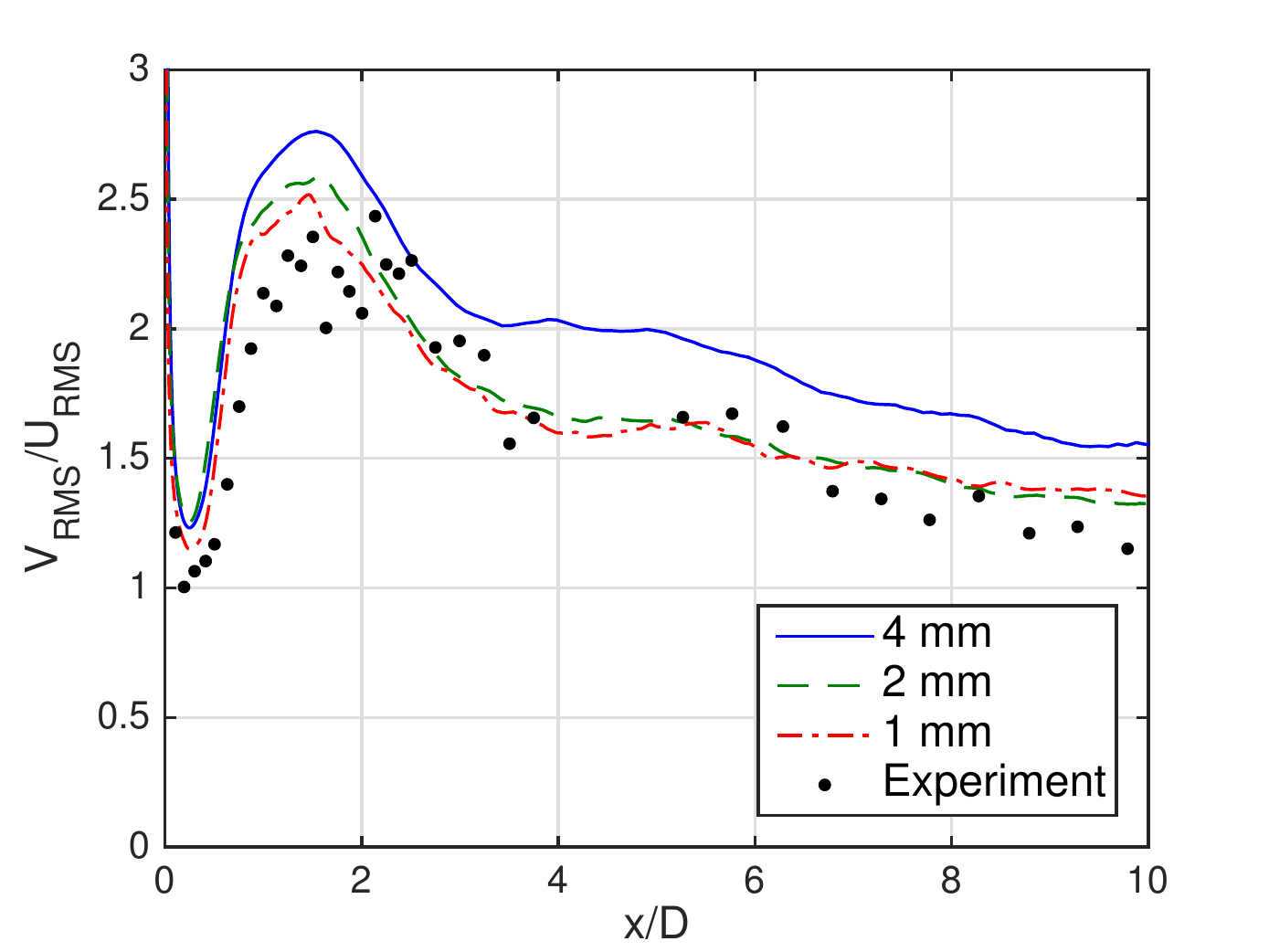}
    }
    \subfigure[Fluctuation]{
        \includegraphics[trim={0 0 0 0},width=0.48\textwidth,clip]{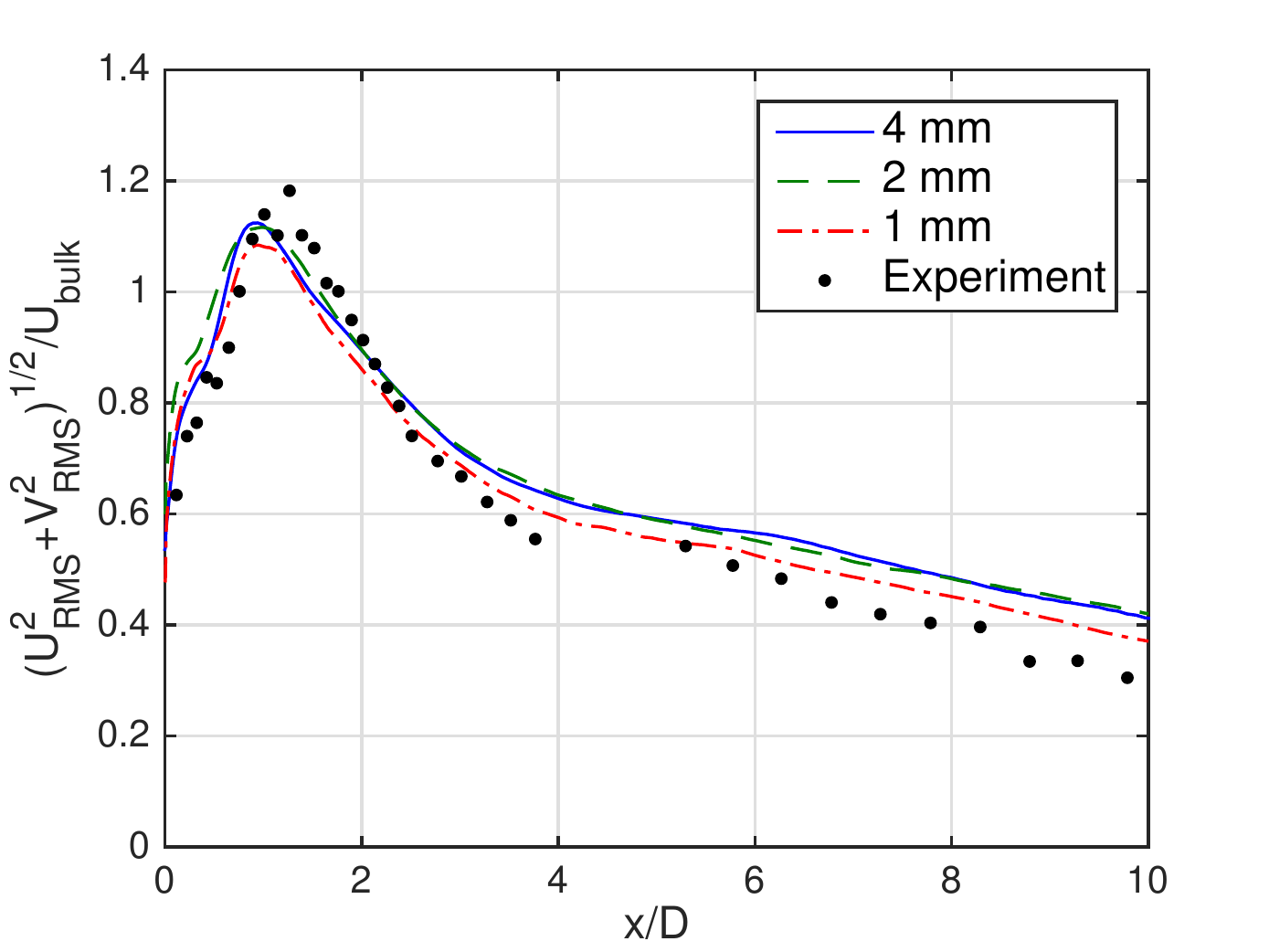}
    }
    \caption{Centerline profiles for (a) mean axial velocity, (b) anisotropy, and (c) fluctuation level on three grids (solid line~\textendash\,~4~mm, dashed line~\textendash\,~2~mm, dash-dotted line~\textendash\,~1~mm) in comparison with measurements (dots) for the non-reacting case from the FV solver. \label{fig:axialprofile}}
\end{figure*}

To further compare the simulation results with the experimental data quantitatively, two more quantities are considered together with the axial velocity at the centerline of the domain downstream the flameholder. The anisotropy and the fluctuation level are defined as $U_\text{RMS} / V_\text{RMS}$ and $\sqrt{U_\text{RMS}^2 + V_\text{RMS}^2} / U_\text{bulk}$, respectively. \Cref{fig:axialprofile} shows the results from simulations using the three grids. It be seen that although the results from three grid resolutions exhibits little difference for the axial velocity and fluctuation level. The 4~mm results for the anisotropy shows behavior quite different to that obtained for the other two grids. This shows that the 2~mm grid resolution is sufficient for the prediction of the instabilities and turbulence for the non-reacting case.

It can be seen from \cref{fig:axialprofile} that, although the anisotropy and fluctuation levels show excellent agreement with the experimental measurements, the axial velocity shows 20\% to 30\% underprediction for axial locations after $x/D = 2$. This discrepancy from the experimental measurements may be due to the different boundary conditions employed in the current simulation to that in the experiments, e.g. the flow is assumed to be laminar at the inlet. Similar underprediction of the axial velocity can be found in numerical simulations using other solvers with similar computational domain and boundary conditions~\cite{cocks2013reacting}. Overall, the non-reacting simulations from the FV solver show very good agreement with the experiments for the available data considered. 

\subsection{Reacting simulations: finite-volume discretization}

\begin{figure*}[!htb!]
    \centering
    \includegraphics[width=0.8\textwidth]{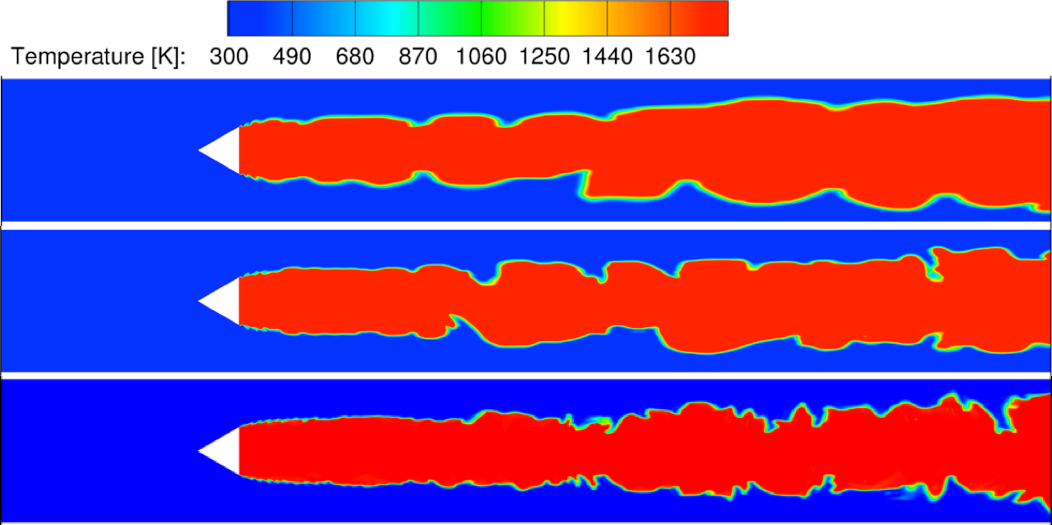}
    \caption{ Instantaneous temperature contours of 4mm (top), 2mm (middle) and 1mm (bottom) cases from the FV solver. \label{fig:inst_temp_R}}
\end{figure*}

\begin{figure*}[!htb!]
    \centering
    \includegraphics[trim={30 0 40 0},width=0.48\textwidth,clip]{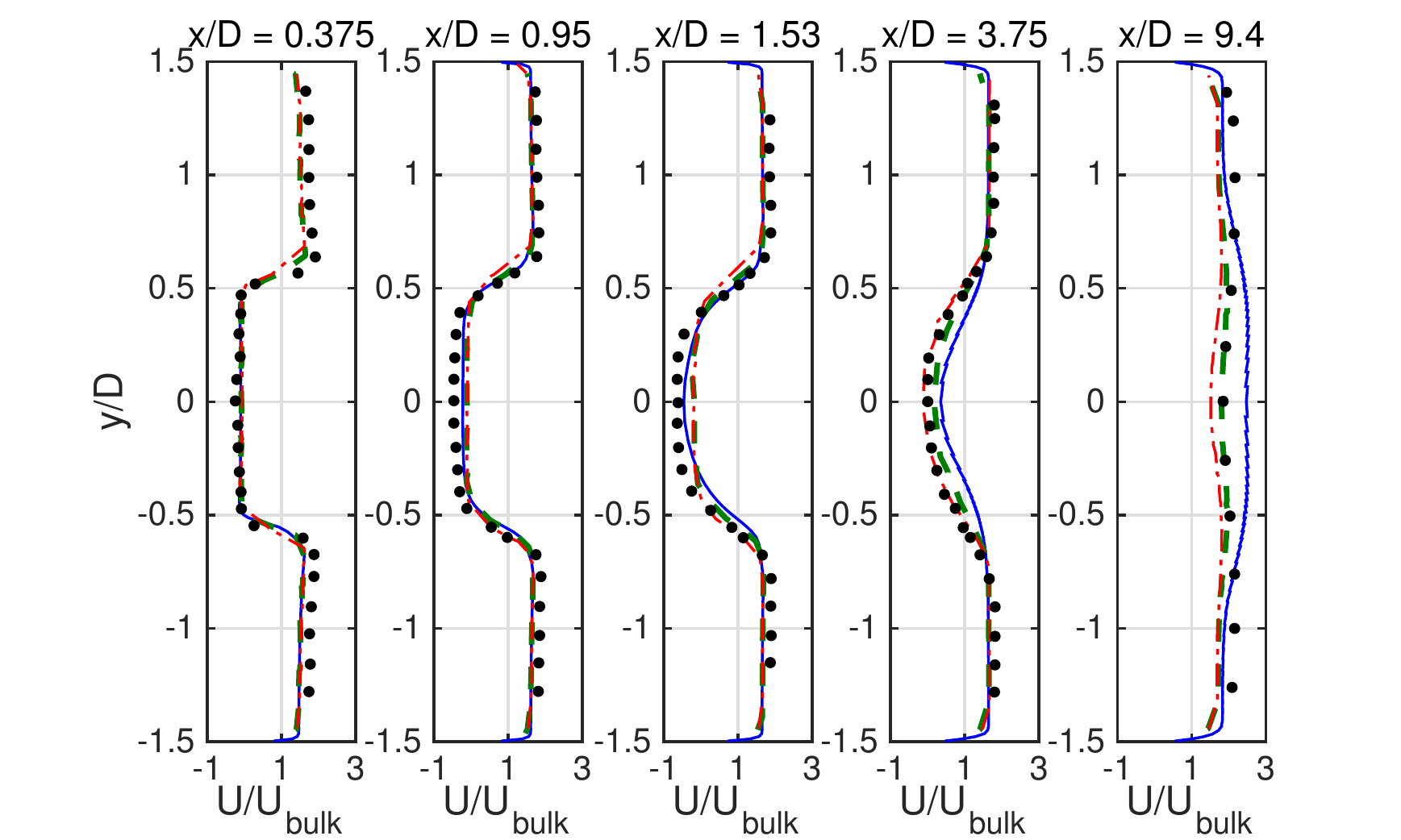}
    \includegraphics[trim={30 0 40 0},width=0.48\textwidth,clip]{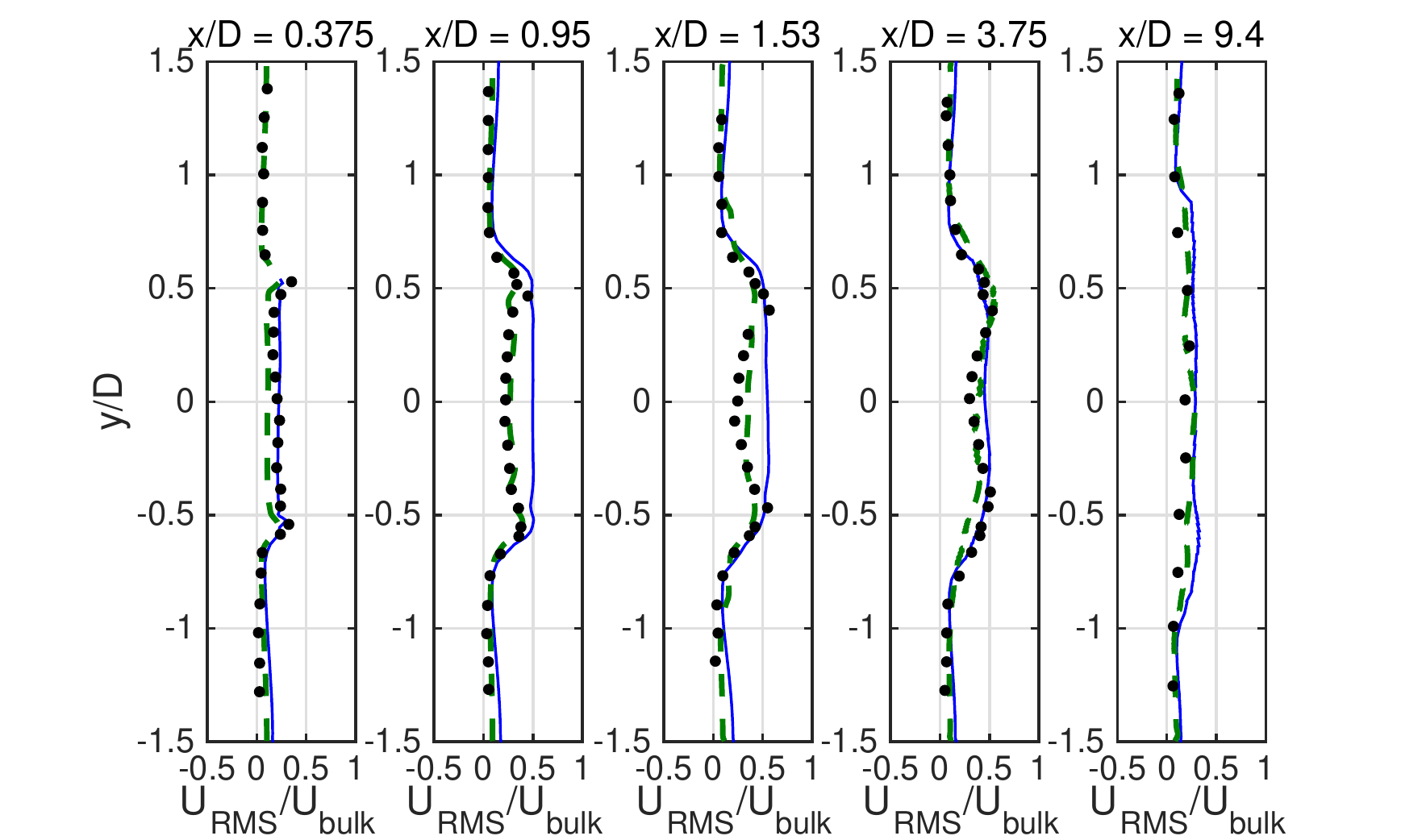}
    \caption{Normalized mean (left) and RMS (right) axial velocity profiles on three grids (solid line~\textendash\,~4~mm, dashed line~\textendash\,~2~mm in comparison with measurements (dots), dash-dotted line~\textendash\,~1~mm), at several axial locations for the non-reacting case from the FV solver. \label{fig:velocityu_R}}
\end{figure*}

\begin{figure*}[!htb!]
    \centering
    \includegraphics[trim={30 0 40 0},width=0.48\textwidth,clip]{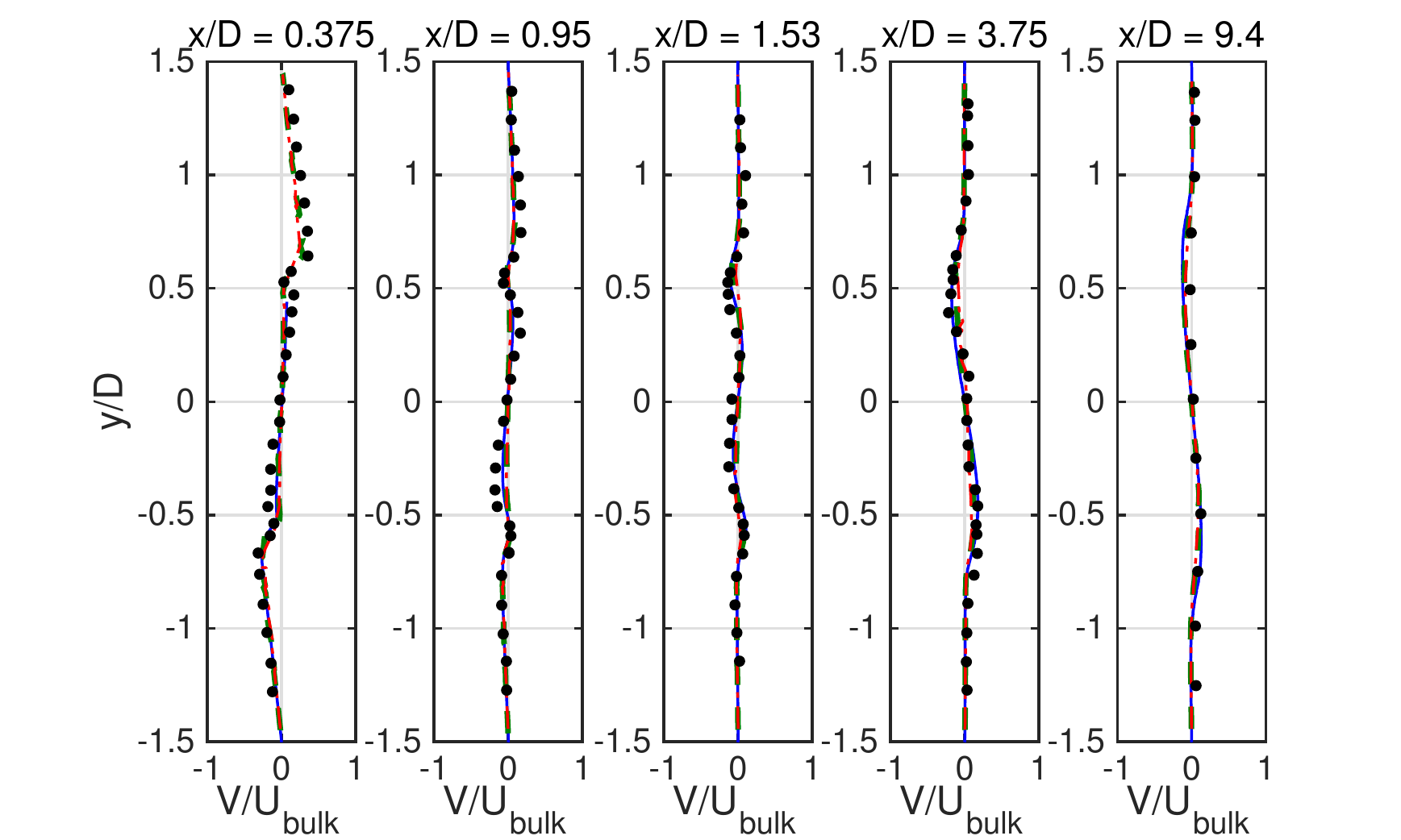}
    \includegraphics[trim={30 0 40 0},width=0.48\textwidth,clip]{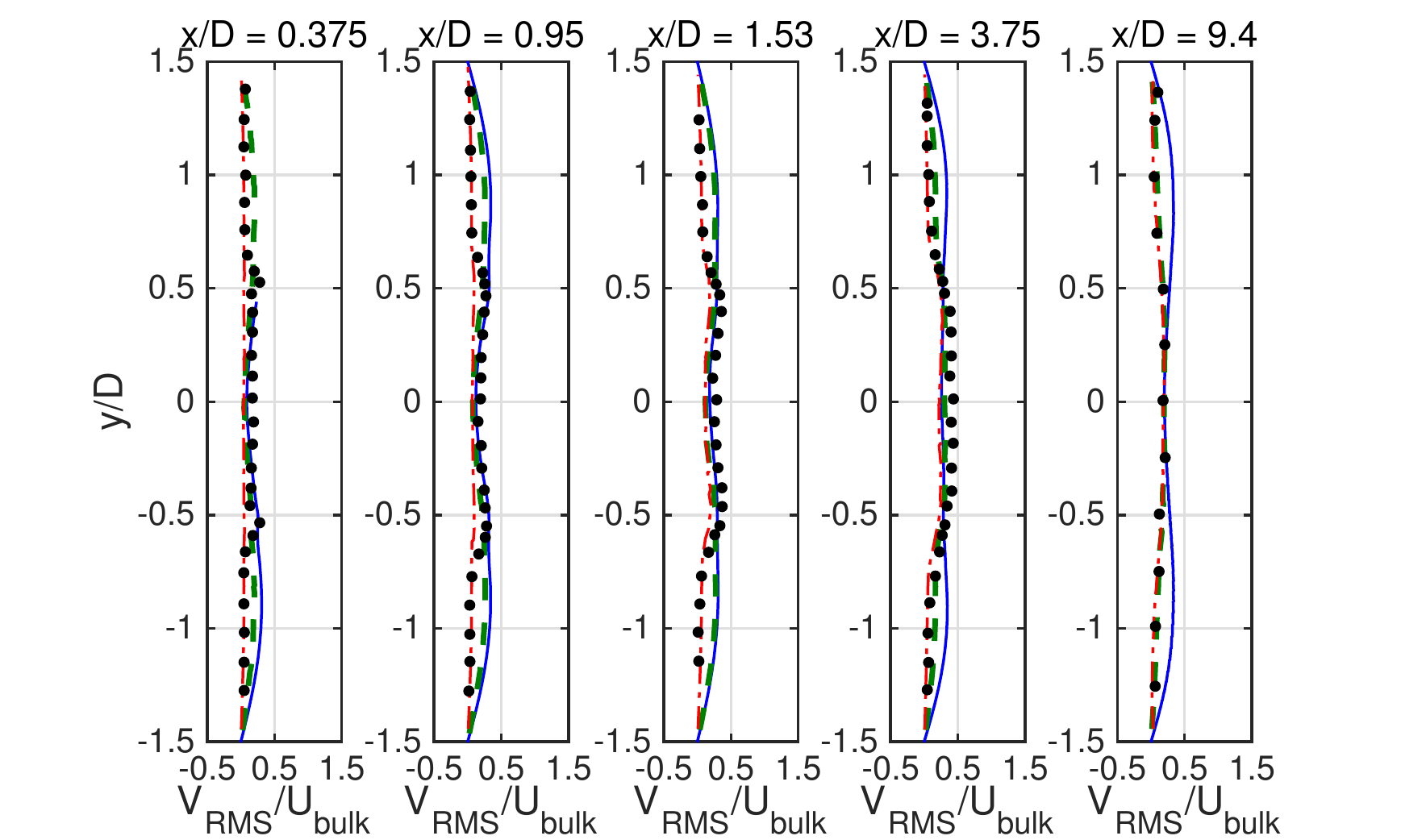}
    \caption{Normalized mean (left) and RMS (right) transverse velocity profiles on three grids (solid line~\textendash\,~4~mm, dashed line~\textendash\,~2~mm, dash-dotted line~\textendash\,~1~mm) in comparison with measurements (dots), at several axial locations for the non-reacting case from the FV solver. \label{fig:velocityv_R}}
\end{figure*}
\begin{figure*}[!htb!]
    \centering
    \includegraphics[trim={30 0 40 0},width=0.48\textwidth,clip]{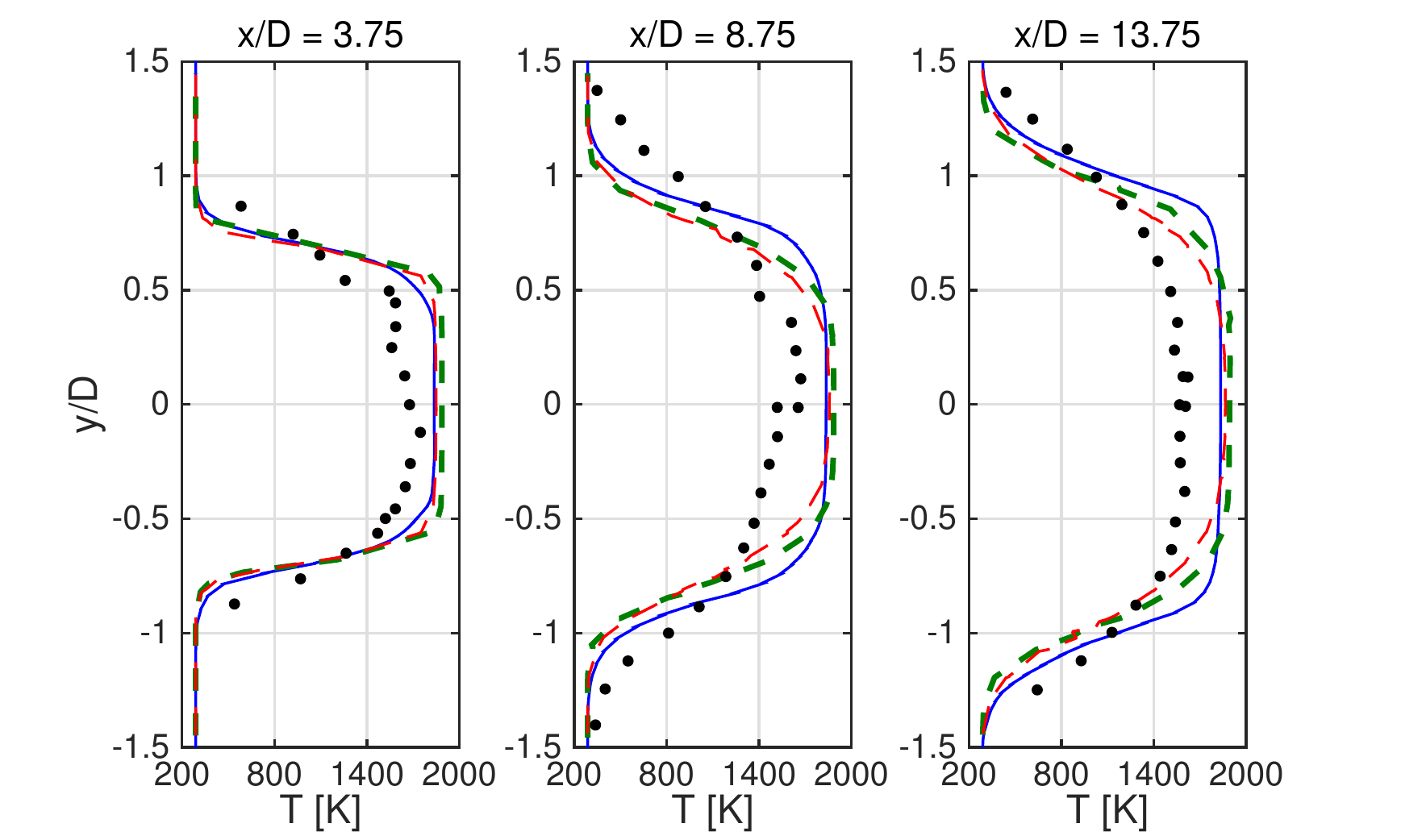}
    \caption{Mean temeparture profiles on three grids (solid line~\textendash\,~4~mm, dashed line~\textendash\,~2~mm, dash-dotted line~\textendash\,~1~mm) in comparison with measurements (dots), at several axial locations for the non-reacting case from the FV solver. \label{fig:temp_R}}
\end{figure*}


The reacting flow simulations with the FV discretization is carried out using the 4mm-resolution, 2mm-resolution, and 1mm-resolution grids. Both simulations use the same combustion model as discussed in Sec.~\ref{SEC_NUM}. 

The elevated temperature in the reacting case as well as the density ratio significantly alter the flow field characteristics. The flow is less turbulent due to the higher viscosity and the large density ratio suppresses sinuous instability mode, which is characteristic for non-reacting flow in the wake of a bluff-body. The laminar flame speed of the mixture with $\phi = 0.65$ is estimated to be $0.2$ m/s and the corresponding flame thickness is $0.6$ mm. Therefore, none of the three FV meshes calculations can resolve the flame and the impact of the flame thickening model on the simulation result is not negligible.

The instantaneous temperature fields of the 4mm and 2mm simulations are shown in Fig.~\ref{fig:inst_temp_R}. Vortex shedding is not present in either simulation. The instability of the shear layer at the near field of the wake is very symmetric. Intermittent sinuous modes can be observed at locations that are further downstream. A stronger effect of flame wrinkling can be observed in the 2mm case, which is not surprising given the higher resolution.

As part of the validation, the mean and RMS axial and transverse velocities across the flame are presented in Fig.~\ref{fig:velocityu_R} and Fig.~\ref{fig:velocityv_R}. Reasonable agreement is obtained by both cases in terms of the mean velocity profiles. The fluctuations are better captured by the higher-resolution case. The achievement of grid convergence cannot be observed from the two cases presented.

\begin{figure*}[!htb!]
    \centering
    \subfigure[Axial velocity]{
        \includegraphics[trim={0 0 0 0},width=0.48\textwidth,clip]{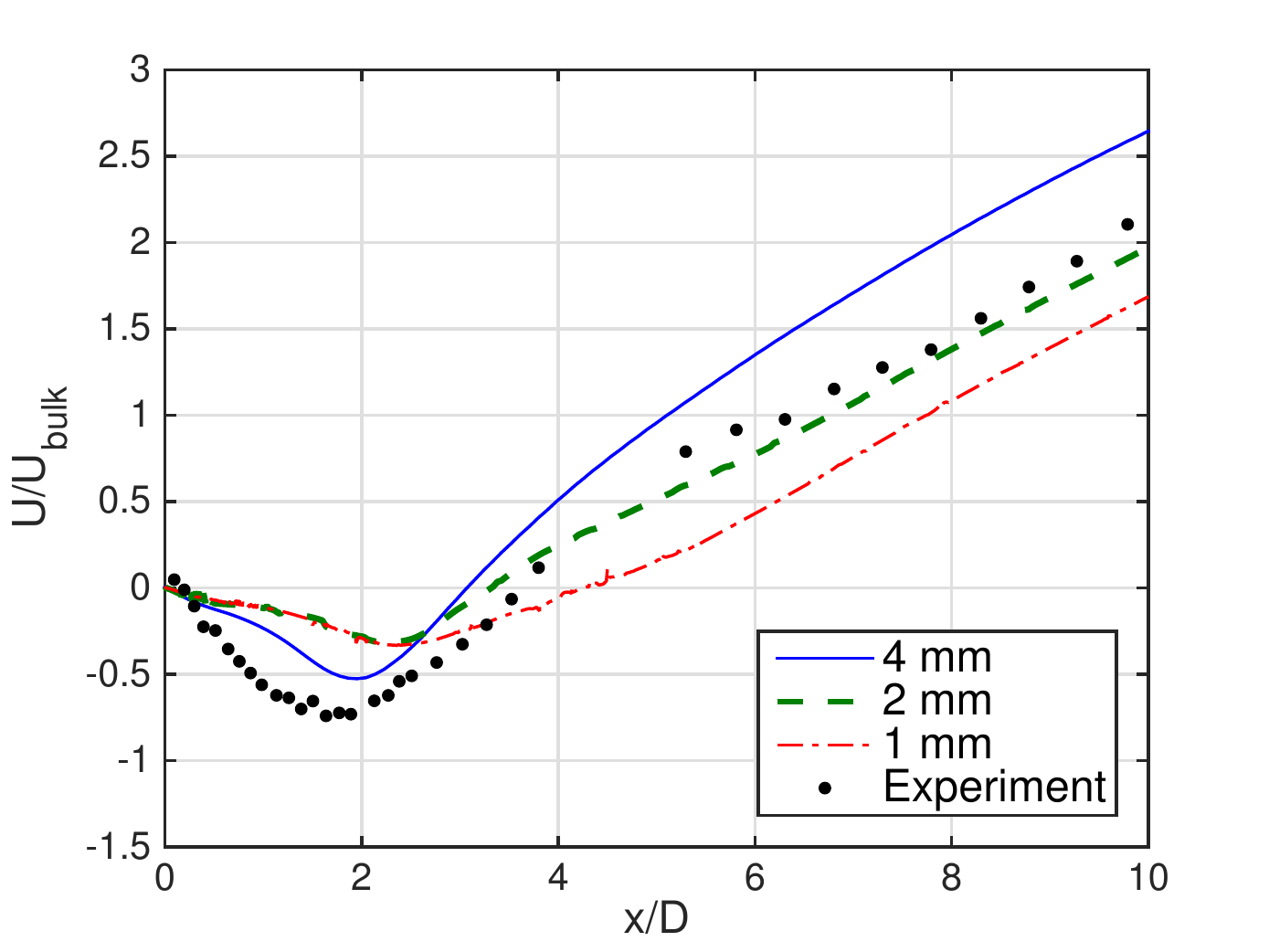}
    }
    \subfigure[Anisotropy]{
        \includegraphics[trim={0 0 0 0},width=0.48\textwidth,clip]{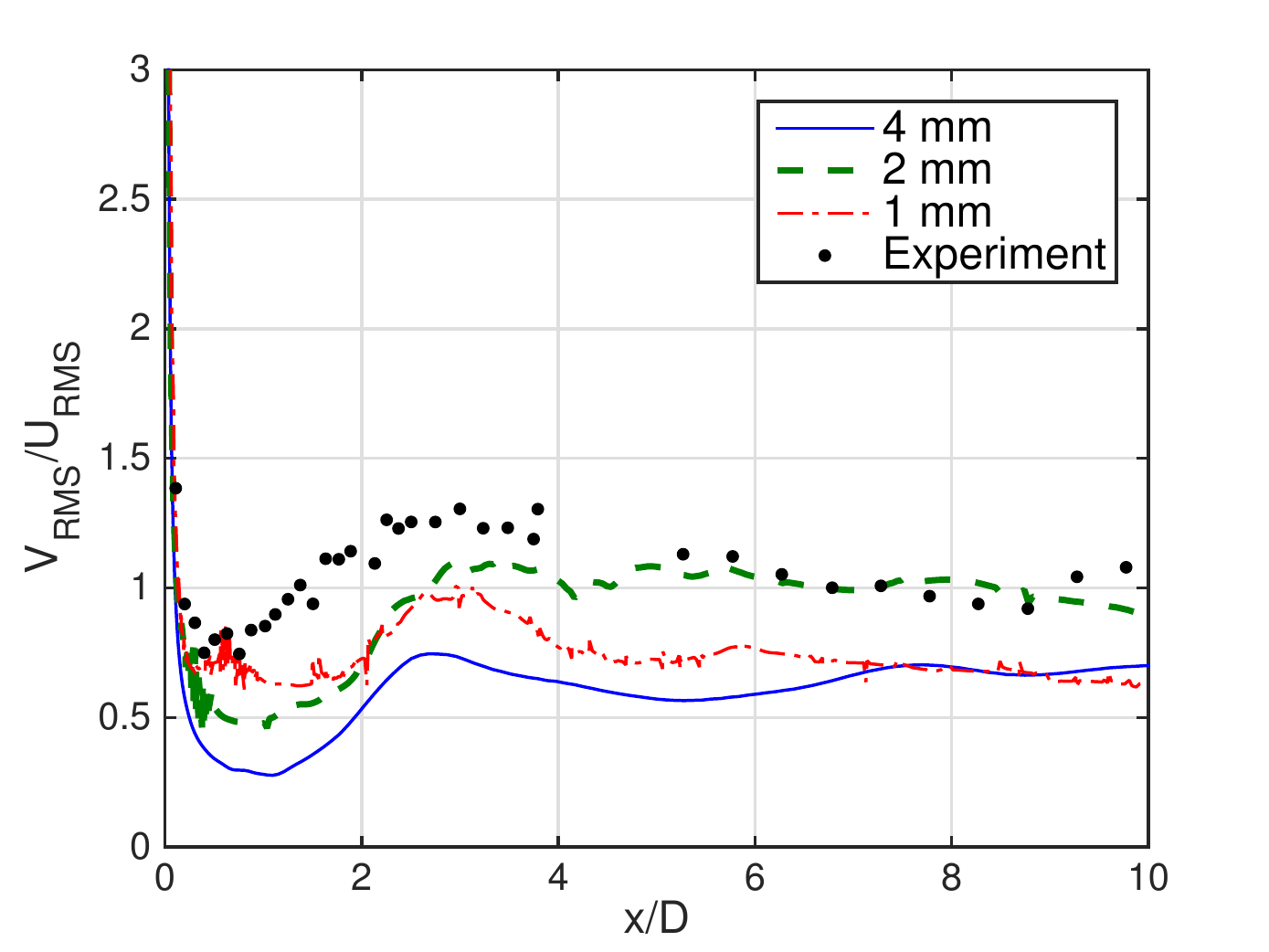}
    }
    \subfigure[Fluctuation]{
        \includegraphics[trim={0 0 0 0},width=0.48\textwidth,clip]{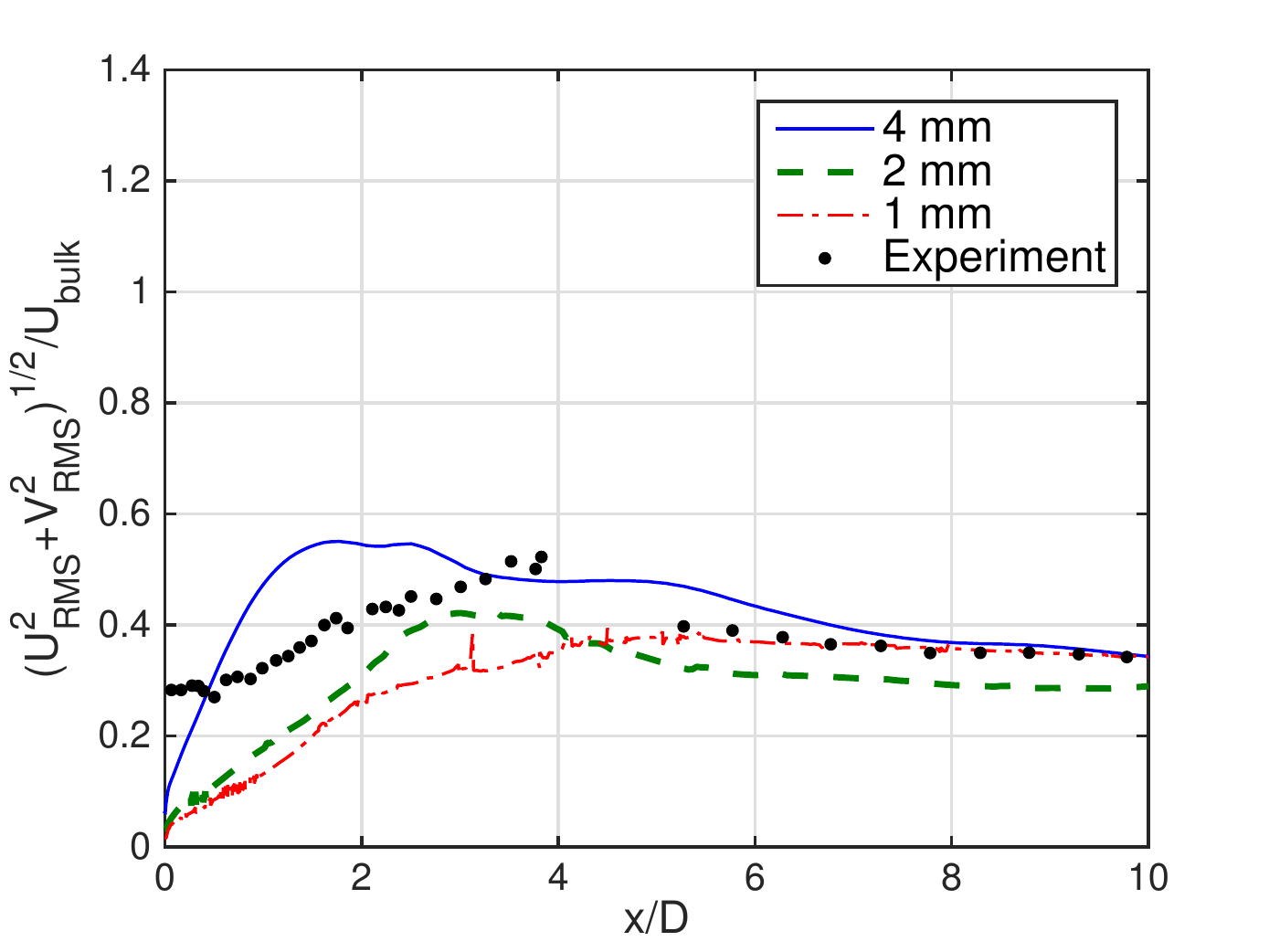}
    }
    \caption{Centerline profiles for (a) mean axial velocity, (b) anisotropy, and (c) fluctuation level on three grids (solid line~\textendash\,~4~mm, dashed line~\textendash\,~2~mm, dash-dotted line~\textendash\,~1~mm) in comparison with measurements (dots) for the non-reacting case from the FV solver. \label{fig:axialprofile_R}}
\end{figure*}

The mean mean axial velocity, anisotropy, and fluctuation along the centerline are shown in Fig~\ref{fig:axialprofile_R}. The cases with finer mesh shows the improvement over all three quantities, but the overall agreement with the experimental data is worse in comparison to the non-reacting case. Grid convergence has not been reached for the series of reacting calculations. The 2~mm case shows overall best agreement with respect to the experimental data among the three cases.

\subsection{Non-reacting simulations: discontinuous-Galerkin discretization}
\begin{figure*}[!htb!]
    \centering
    \subfigure[Mean axial velocity]{\includegraphics[trim={38 0 25 0},width=0.48\textwidth,clip]{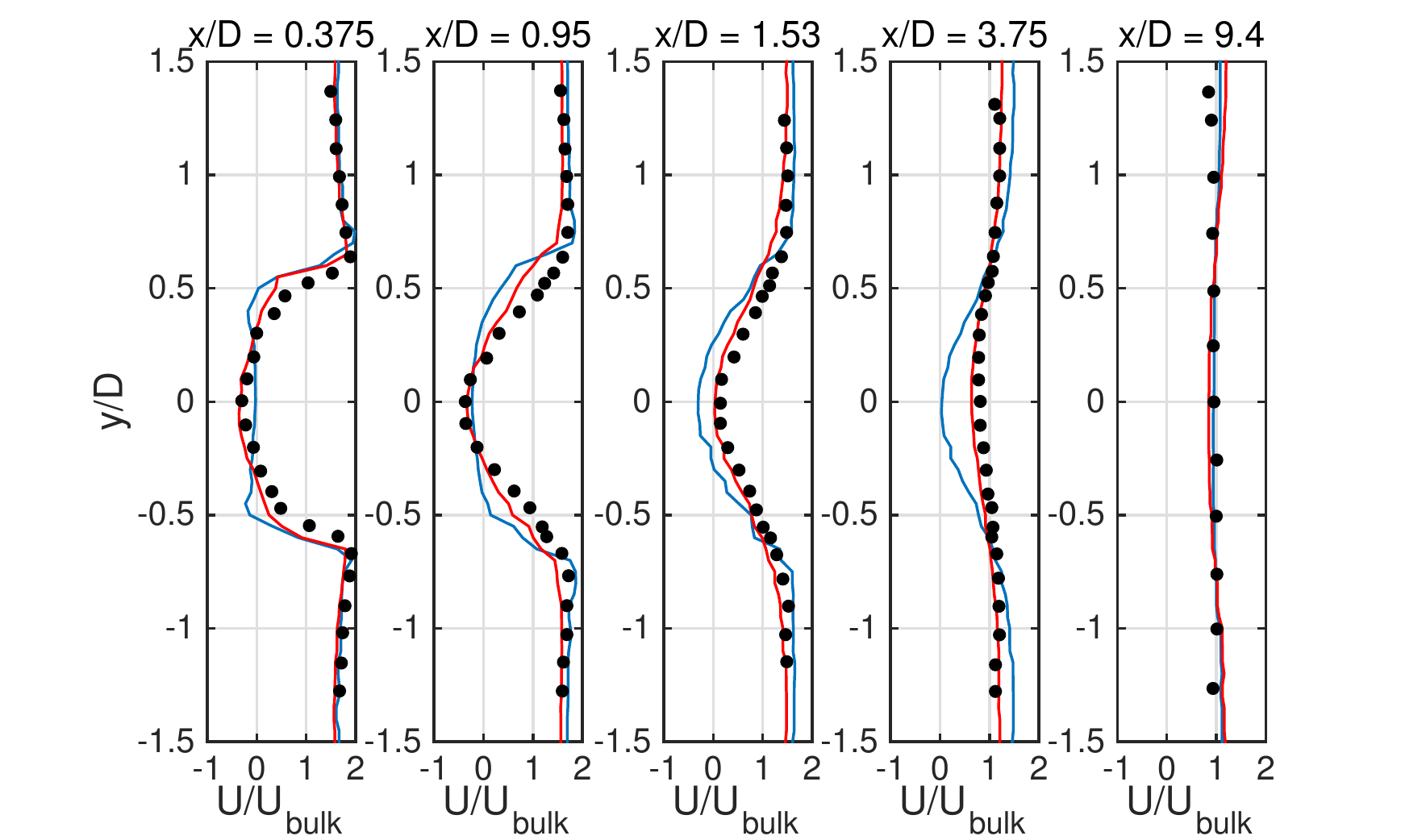}\label{RESULT_NR_DG_a}}
    \subfigure[RMS axial velocity]{\includegraphics[trim={38 0 25 0},width=0.48\textwidth,clip]{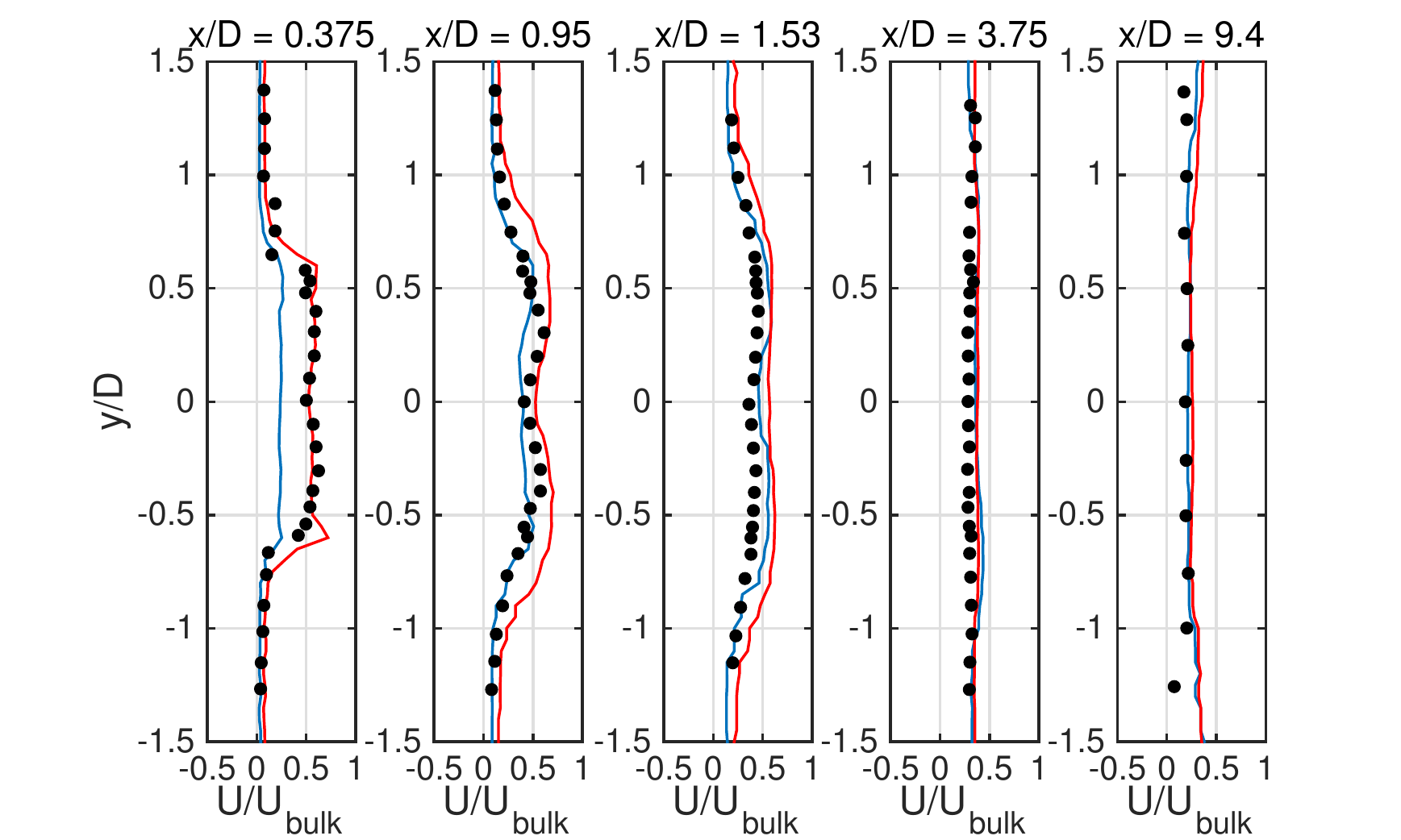}\label{RESULT_NR_DG_b}}
    \subfigure[Mean transverse velocity]{\includegraphics[trim={38 0 25 0},width=0.48\textwidth,clip]{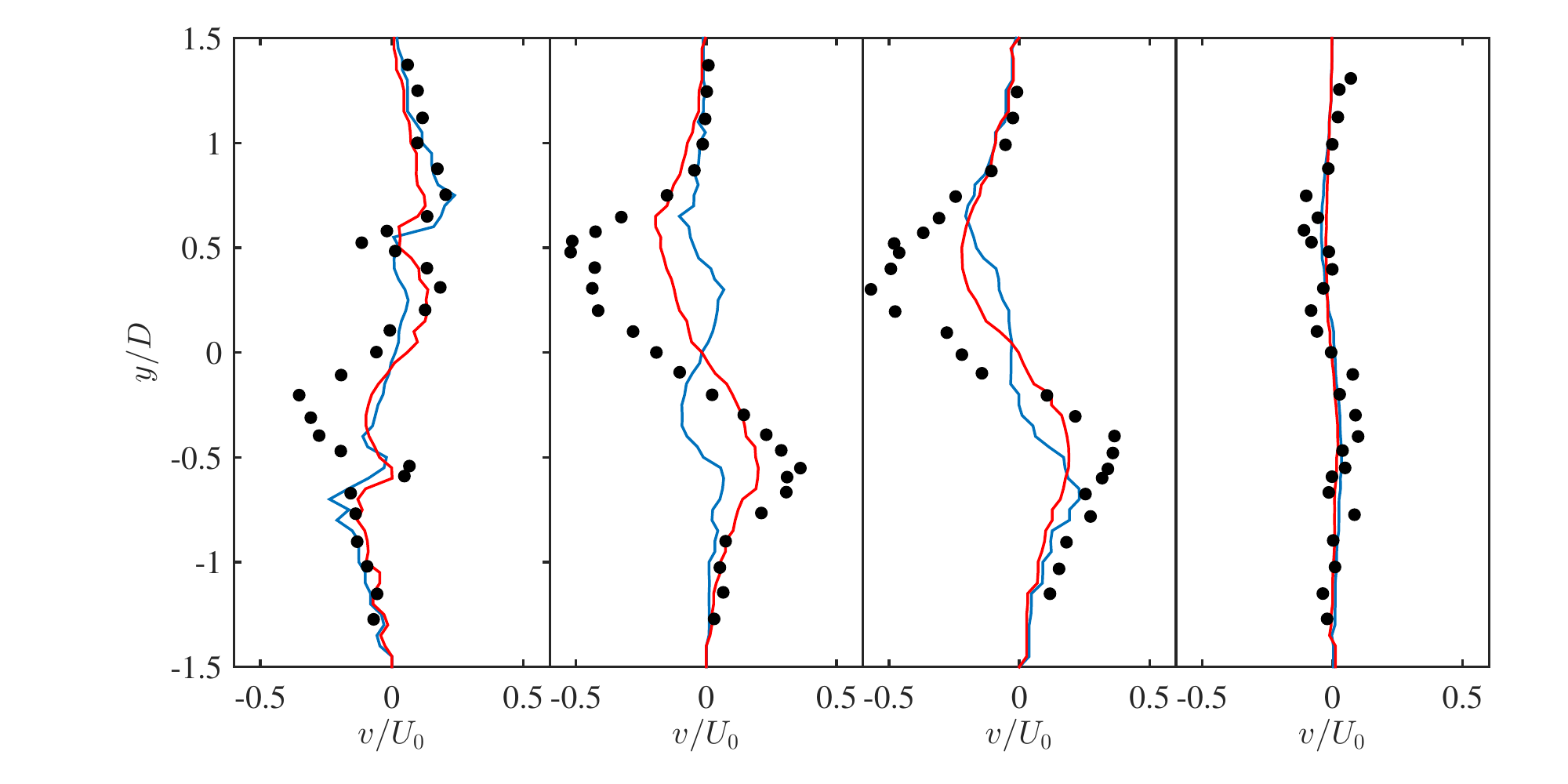}\label{RESULT_NR_DG_c}}
    \subfigure[RMS transverse velocity]{\includegraphics[trim={38 0 25 0},width=0.48\textwidth,clip]{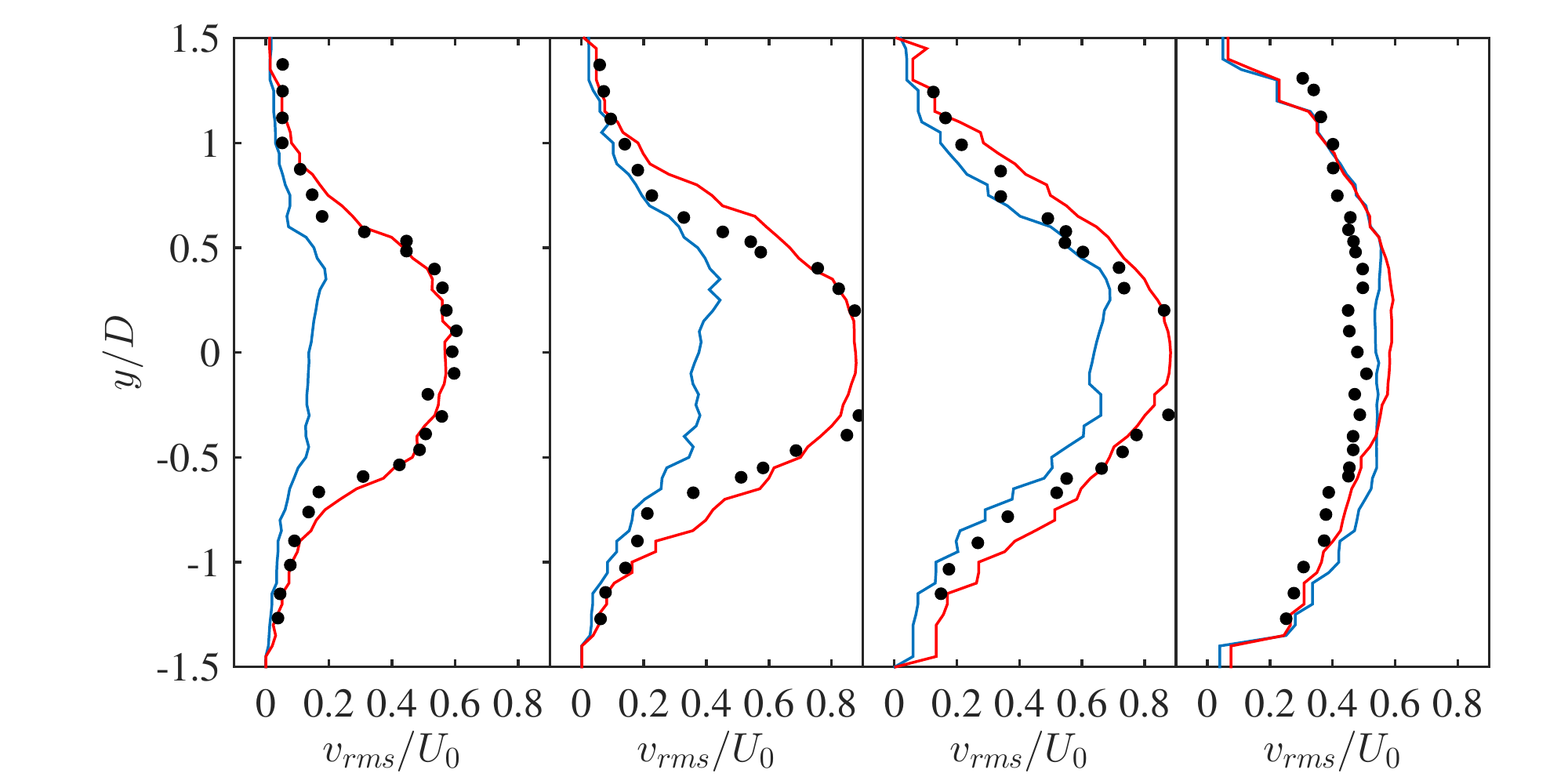}\label{RESULT_NR_DG_d}}
    \caption{\label{RESULT_NR_DG}DG simulation results for the non-reacting case (blue line $-$ DGP1; red line $-$ DGP2; dots $-$ experimental data).}
\end{figure*}

\Cref{RESULT_NR_DG} shows DG simulation results for the non-reacting case. The axial velocity profile shows a noticeable recirculation zone behind the bluff body. The DG solutions capture the mean-flow profiles along the transverse direction, especially for the DGP2 solution. However, DGP1 method predicts a larger recirculation zone, showing considerable discrepancy from the measurement. As for the profile of the transverse velocity, both DGP1 and DGP2 solutions shows insufficient agreement with experimental data. That is likely due to the simplification made on inflow condition (no consideration of inflow turbulence) and the size reduction of the domain along the spanwise direction. Comparatively speaking, DGP2 solution is still able to provide a better representation of the variation of the transverse velocity along wall-normal direction, compared to the DGP1 solution. Velocity RMS predictions are presented in \cref{RESULT_NR_DG_b,RESULT_NR_DG_d}. Both DGP1 and DGP2 simulations are able to provide reasonable representation of the RMS of streamwise velocity. At the upstream location very close to the bluff body, DGP2 solution shows considerably better agreement with the measured RMS of streamwise velocity, compared to the DGP1 solution. As for the RMS of the transverse velocity, DGP2 prediction is consistently better than that of DGP1 at all measurement locations. 
 
\subsection{Reacting simulations: discontinuous-Galerkin discretization}
\begin{figure*}[!htb!]
    \centering
    \subfigure[Mean axial velocity]{\includegraphics[trim={40 0 25 0},width=0.48\textwidth,clip]{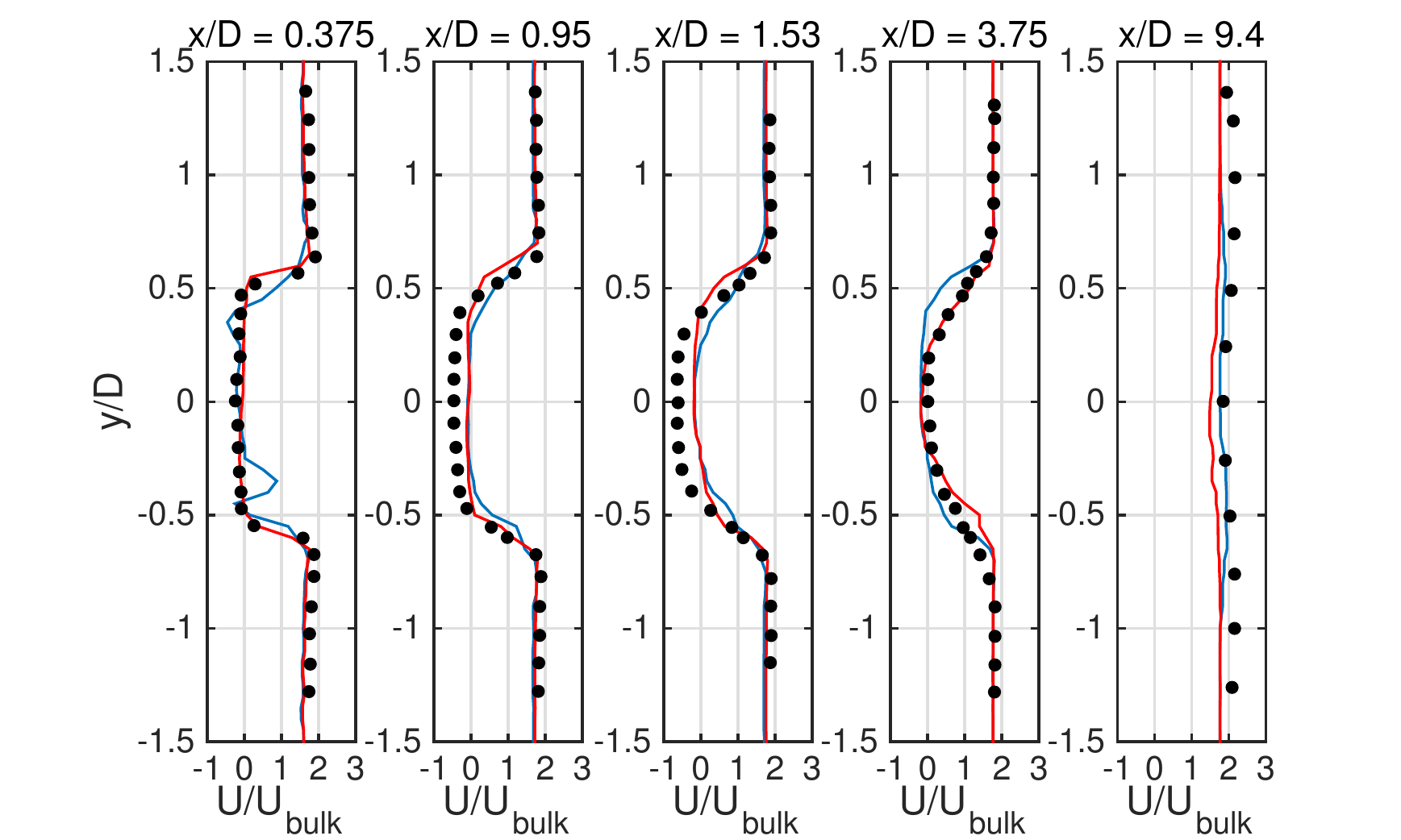}}
    \subfigure[Mean transverse velocity]{\includegraphics[trim={40 0 25 0},width=0.48\textwidth,clip]{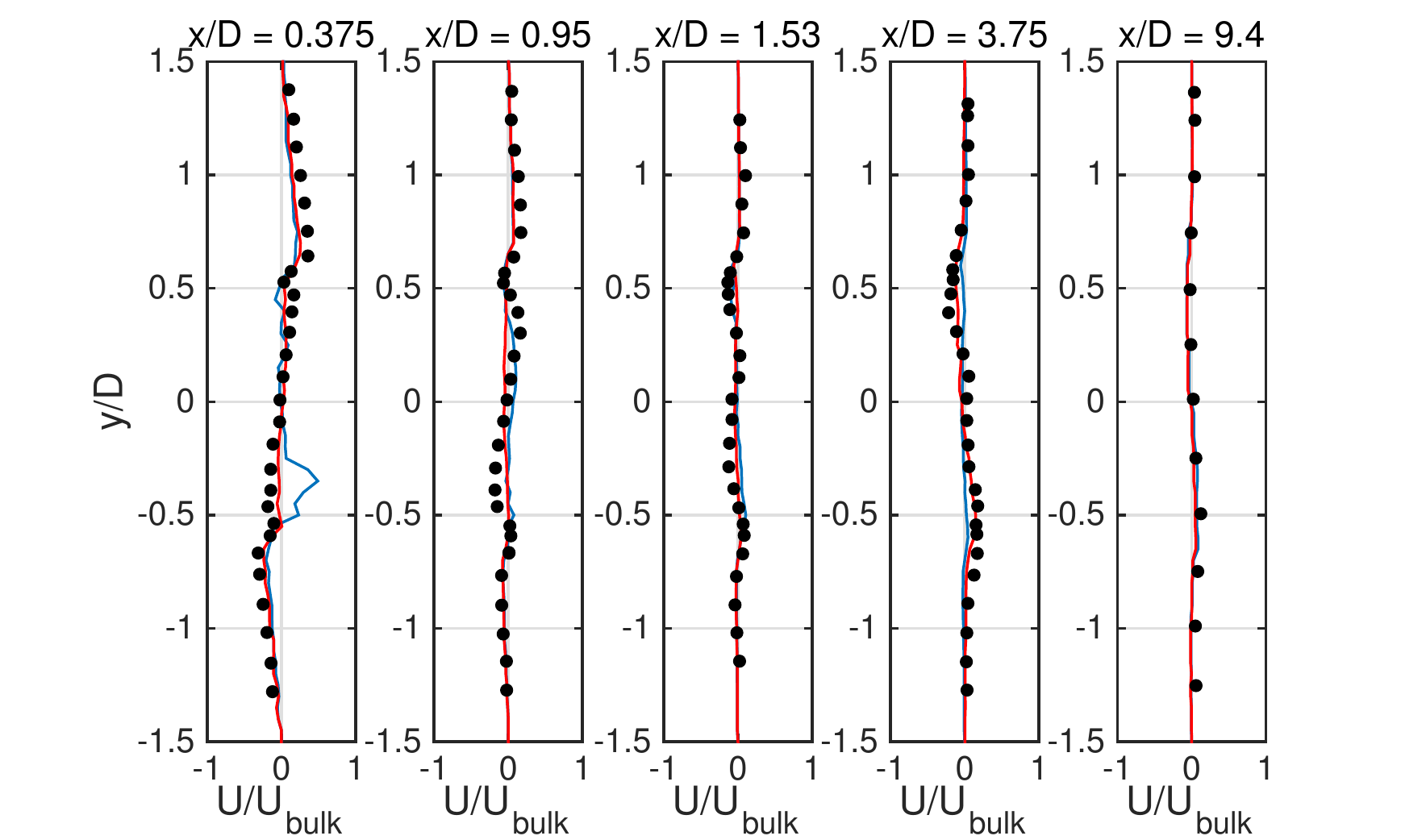}}
    \subfigure[Temperature]{\includegraphics[trim={40 0 25 0},width=0.48\textwidth,clip]{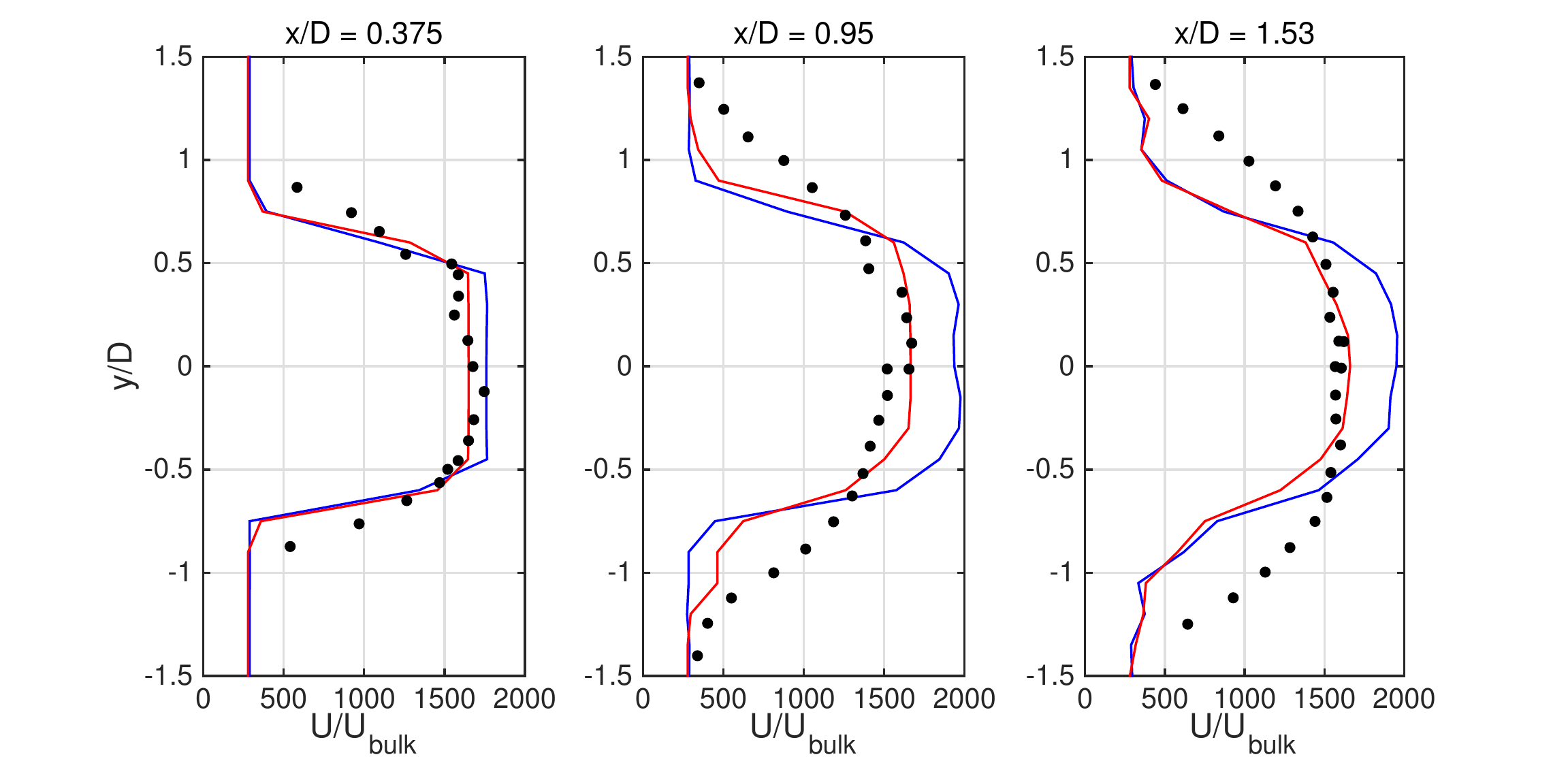}\label{RESULT_R_DG_c}}
    \caption{\label{RESULT_R_DG}DG simulation results of the mean-flow profile for the reacting case (blue line $-$ DGP1; red line $-$ DGP2; dots $-$ experimental data).}
\end{figure*}

\Cref{RESULT_R_DG} shows DG prediction of the mean-flow profile for the reacting case. Compared to the non-reacting case in \cref{RESULT_NR_DG_a}, sharper gradients appear in the axial velocity profile between the forward and backward flows due to the presence of the premixed flame. Both DGP1 and DGP2 solutions show reasonably good agreement with the experimental data. Comparatively, DGP2 solution is superior in terms of representing near-body flow profiles and preserving the velocity gradient at downstream locations. Both DGP1 and DGP2 methods, however, predict a slightly weaker back-flow in the recirculation zone compared to the experimental data. As for the prediction of the transverse velocity, results from DGP1 and DGP2 show difficulties near the bluff-body. DGP1 solution has large discrepancy from the measurement. This problem is likely due to the lack of numerical resolution to represent the interaction between the flame and the flow field. The lack of resolution in the DGP1 simulation might lead to a poor representation of chemical source terms, which could explain the consistent over-prediction of the centerline temperature. In contrary, DGP2 simulation can accurately predict the equilibrium temperature along the centerline. The measurement shows a thick flame brush compared to the numerical predictions by the DG methods. 

\section{Conclusions} \label{SEC_CONC}
A series of LES simulations were performed for the Volvo bluff-body stabilized combustion configuration using two solvers featuring a high-resolution finite-volume method as well as a high-order DG discretization. Both the non-reacting and reacting cases are calculated. For the non-reacting cases, good agreement with the experimental data is achieved by solvers at high numerical resolution (2mm case for FV and DGP2 for DG). The reacting cases are more challenging due to the small length scale of the flame and the suppression of sinuous mode of absolute instability by the density ratio. Improvements over the results of higher-resolution cases were observed for both solvers. Grid convergence is not achieved by either solver. The impact of the models for turbulence-combustion interaction is important at the current resolutions for both solvers. A closer investigation of this aspect is the subject of future research.

\section*{Acknowledgments}
This work was funded by NASA Transformational Tools and Technologies Project with Award No. NNX15AV04A.

\bibliography{bibliography,bibliography_DG}
\bibliographystyle{aiaa}

\end{document}